\documentclass[journal, twoside, final]{IEEEtran}
\usepackage{amsmath, amssymb, amsfonts} 
\usepackage{algorithmic}
\usepackage{array}
\usepackage[caption=false, font=scriptsize, labelfont=rm, textfont=rm]{subfig}
\usepackage{textcomp}
\usepackage{stfloats}
\usepackage{url}
\usepackage{verbatim}
\usepackage{graphicx}
\usepackage{cite}

\usepackage{bm, color}
\usepackage{booktabs}
\usepackage{setspace}
\usepackage{longtable}
\usepackage{makecell}

\usepackage{dsfont}
\usepackage{datetime}
\usepackage{float}
\usepackage{bbm}
\usepackage{siunitx, nicematrix}
\usepackage{enumitem}
\usepackage{nicematrix}
\usepackage{multicol}
\usepackage{longtable, threeparttable, multirow}

\makeatletter
\newcommand\figcaption{\def\@captype{figure}\caption}
\newcommand\tabcaption{\def\@captype{table}\caption}
\makeatother

\newtheorem{assumption}{Assumption}

\newtheorem{theorem}{Theorem}
\newtheorem{remark}{Remark}

\usepackage[ruled]{algorithm2e}

\allowdisplaybreaks[4]

\begin{document}

\title{\huge Air-to-Ground Communications for Internet of Things: UAV-based Coverage Hole Detection and Recovery} 
\author{Xiao Fan, Wenkun Wen,~\IEEEmembership{Member,~IEEE}, Peiran Wu,~\IEEEmembership{Member,~IEEE}, Junhui Zhao, \IEEEmembership{Senior~Member,~IEEE}, and Minghua Xia, \IEEEmembership{Senior~Member,~IEEE}
	\thanks{Received 12 October 2025; revised 11 Decemeber 2025 and 02 Janunary 2025; accepted 07 Janunary 2025. \textit{(Corresponding authors: Wenkun Wen; Minghua Xia)}}
	\thanks{Xiao Fan was with the School of Electronics and Information Technology, Sun Yat-sen University, Guangzhou 510006, China; she is now with the School of Information and Communication, Guilin University of Electronic Technology, Guilin 541004, China (e-mail: fanx26@mail2.sysu.edu.cn).}
	\thanks{Wenkun Wen is with the R\&D Department of the Techphant Technologies Company Ltd., Guangzhou 510310, China (email: wenwenkun@techphant.net).}
	\thanks{Peiran Wu, and Minghua Xia are with the School of Electronics and Information Technology, Sun Yat-sen University, Guangzhou 510006, China (e-mail: wupr3@mail.sysu.edu.cn, xiamingh@mail.sysu.edu.cn).} 
	\thanks{Junhui Zhao is with the School of Electronic and Information Engineering, Beijing Jiaotong University, Beijing 100044, China (email: junhuizhao@bjtu.edu.cn).}
	\thanks{Color versions of one or more of the figures in this article are available online at https://ieeexplore.ieee.org.}	
	\thanks{Digital Object Identifier }
}

\markboth{IEEE Internet of Things Journal} {Fan \MakeLowercase{\textit{et al.}}: Air-to-Ground Communications for Internet of Things: UAV-based Coverage Hole Detection and Recovery}

\maketitle

\IEEEpubid{\begin{minipage}{\textwidth} \ \\[12pt] \centering 2327-4662 \copyright\ 2026 IEEE. All rights reserved, including rights for text and data mining, and training of artificial intelligence \\ and similar technologies. Personal use is permitted, but republication/redistribution requires IEEE permission. \\
See \url{https://www.ieee.org/publications/rights/index.html} for more information.\end{minipage}}

 \IEEEpubidadjcol

\begin{abstract}
    \noindent  Uncrewed aerial vehicles (UAVs) play a pivotal role in ensuring seamless connectivity for Internet of Things (IoT) devices, particularly in scenarios where conventional terrestrial networks are constrained or temporarily unavailable. However, traditional coverage-hole detection approaches, such as minimizing drive tests, are costly, time-consuming, and reliant on outdated radio-environment data, making them unsuitable for real-time applications. To address these limitations, this paper proposes a UAV-assisted framework for real-time detection and recovery of coverage holes in IoT networks. In the proposed scheme, a patrol UAV is first dispatched to identify coverage holes in regions where the operational status of terrestrial base stations (BSs) is uncertain. Once a coverage hole is detected, one or more UAVs acting as aerial BSs are deployed by a satellite or nearby operational BSs to restore connectivity. The UAV swarm is organized based on Delaunay triangulation, enabling scalable deployment and tractable analytical characterization using stochastic geometry. Moreover, a collision-avoidance mechanism grounded in multi-agent system theory ensures safe and coordinated motion among multiple UAVs. Simulation results demonstrate that the proposed framework achieves high efficiency in both coverage-hole detection and on-demand connectivity restoration while significantly reducing operational cost and time.
\end{abstract}

\begin{IEEEkeywords}
	Air-ground communications, collision avoidance, coverage hole, Delaunay triangulation, uncrewed aerial vehicle (UAV). 
\end{IEEEkeywords}

\section{Introduction} \label{sec: introduction}    
\IEEEPARstart{T}{o} support  diverse use cases and applications ubiquitously, forthcoming sixth-generation (6G) wireless networks must deliver wide-area coverage with high capacity and ultra-low latency \cite{Jiang2021Road, Wang2023Road}. In the context of the Internet of Things (IoT), seamless communication coverage is particularly crucial for ensuring reliable data exchange among interconnected devices \cite{Cheng2023AI}. To meet the stringent connectivity requirements of IoT applications, telecom operators must eliminate coverage holes and mitigate poor service quality across cellular networks. In practical deployments, a \emph{coverage hole} refers to a region where the received signal strength from the serving base station (BS) or its neighboring cooperative BSs falls below the threshold required to sustain the minimum quality of service or acceptable radio link performance \cite{Gomez-Andrades2016Method}. When IoT devices or user equipment (UE) enter such regions, they may experience communication disruption, degraded service quality, or even radio link failure, which can critically impair IoT operations. Coverage holes may arise from various causes, including physical obstructions, suboptimal antenna configurations, hardware malfunctions, improper frequency planning, or natural disasters \cite{3GPPTS37.320}. Furthermore, millimeter-wave (mmWave) communications—a key enabler of 6G networks—are particularly susceptible to signal blockage and attenuation, thereby increasing the likelihood of coverage holes \cite{Anjinappa2021Coverage}.
  
    \subsection{Related Works and Motivation} \label{sec:introduction-motivation}
        Conventional approaches for detecting coverage holes in legacy cellular networks rely on a combination of drive tests, customer complaints, and software or hardware alarms \cite{Akbari2016How}. These methods are typically costly, time-consuming, and prone to inaccuracies. To alleviate these limitations, the minimization of drive tests (MDT) technique was standardized by the 3rd Generation Partnership Project (3GPP), enabling operators to automatically collect user measurements and signaling messages \cite{Hapsari2012Minimization}. However, due to their dependence on user equipment (UEs), MDT-based approaches suffer from several drawbacks: UEs cannot always provide immediate reports, and their location information may be imprecise or unavailable due to privacy constraints \cite{Akbari2016How}. Consequently, such methods are effective mainly in urban or suburban environments but largely impractical in rural regions or post-disaster scenarios (e.g., earthquakes or hurricanes), where network infrastructure may be damaged, and the operational status of terrestrial base stations (BSs) is unknown.

 \IEEEpubidadjcol
 
        An emerging solution to these challenges is the adoption of mobile robotic platforms, such as uncrewed aerial vehicles (UAVs) \cite{Mozaffari2019Tutorial}. With the aid of satellites and UAV-based aerial base stations (ABSs), traditional wireless networks can be extended into three-dimensional (3D) space to achieve ubiquitous broadband connectivity \cite{Qi2023Key}. Compared with satellites, UAVs offer greater flexibility and cost efficiency, making them suitable for providing temporary wireless coverage in inaccessible areas such as caves or tunnels. In emergencies caused by natural disasters, UAVs can rapidly reestablish communication links using their strong line-of-sight (LoS) channels \cite{Liu2019DSF-NOMA}. High-speed backhaul connections to UAVs can be supported by nearby BSs \cite{Li2023} or satellites \cite{Hu2020Joint}.

        Beyond single-UAV deployments, an \emph{UAV swarm} comprising multiple UAVs can perform complex tasks more efficiently by collectively gathering richer spatiotemporal information. Coordinated multi-point (CoMP) transmission techniques \cite{3GPPTR36.819} can be further integrated to enhance UAV-enabled communications. To systematically develop such swarms, Delaunay triangulation---a mathematically tractable geometric structure widely applied in stochastic geometry---can be exploited to form CoMP clusters for both air-to-air \cite{Li2020B} and air-to-ground\cite{Fan2023} networks.

        Effective movement control is crucial for exploiting UAV swarm mobility in wireless networks. Existing research on UAV movement control has primarily focused on robotics and control theory \cite{Qian2021Formation}, aiming to design strategies that drive UAVs to form specific geometric patterns and maneuver cohesively \cite{Shakeri2019Design}. Depending on the control paradigm, UAV formation control can be broadly categorized as leader-follower or leaderless \cite{Oh2015Survey}. For terrestrial coverage-hole recovery, leaderless structures are particularly advantageous, as all UAVs share identical functionality, enhancing robustness against individual UAV failures.

        In real-world environments, collisions between UAVs or with external obstacles remain a significant risk \cite{Yasin2020UAV}. To enhance operational reliability, various collision-avoidance schemes have been proposed \cite{Senanayake2016Search}. Among them, the \emph{potential function-based method} is widely used for multi-agent systems due to its smooth trajectory generation and simplicity of implementation \cite{Zhu2019flexible}. The technique is inspired by the attractive and repulsive forces among charged particles \cite{Khatib1985}. Recent studies have combined the potential field approach with error transformation to achieve collision-free and connectivity-preserving formations \cite{Yang2023Collision}, and have integrated potential functions with anti-windup compensators to construct adaptive, saturated formation-control schemes \cite{Lu2022Collision}. 
        
        Beyond classical collision-avoidance strategies, recent research has increasingly focused on intelligent UAV swarm control using deep reinforcement learning and neural network-based methods. For example, \cite{Marek2025A2G} reviews collision-avoidance mechanisms for swarms of drones, while deep reinforcement learning has been applied to formation control with collision avoidance~\cite{Sui2021Formation}, and to more general collision-free motion planning in cluttered environments, such as the URPlanner framework~\cite{Ying2025URPlanner}; neural-network-based formation control with collision/obstacle avoidance and connectivity maintenance has also been investigated in~\cite{Aryankia2022NN}. These works highlight a promising direction towards more autonomous and adaptive swarm behaviors. In contrast, the present paper focuses on potential-function-based smooth trajectory generation, which can serve as a complementary, model-based component or safety layer for future intelligent cooperative control frameworks.

        Motivated by these insights, this paper employs the regular tetrahedron---the optimal configuration of Delaunay triangulation in 3D space---to model the formation of the UAV swarm. This geometric structure offers two main advantages: {\it (i)} its symmetry and simplicity facilitate formation maintenance and scalable swarm integration, and {\it (ii)} it can be extended to 3D air-to-air networks to enable autonomous self-healing of aerial infrastructures, a key component of future space–air–ground integrated networks \cite{Li2020B}. Building upon this framework, two UAV scheduling algorithms are developed based on the detected coverage hole size. Finally, a multi-UAV movement-control strategy with collision avoidance is designed using the Lyapunov stability theory \cite{Cao2013Overview}, ensuring global asymptotic stability of the UAV swarm.
    
    \subsection{Summary of Major Contributions} \label{sec:introduction-contribution}
        This paper develops a UAV-assisted framework for detecting and recovering terrestrial coverage holes in large-scale IoT networks. The framework integrates network modeling, UAV scheduling, and multi-UAV control with collision avoidance. The main contributions are summarized as follows:
		\begin{enumerate}[label={\arabic*)}]
		    \item Network modeling: A novel air-to-ground wireless network model is proposed for terrestrial coverage-hole detection and recovery, applicable to both emergency scenarios (e.g., earthquakes and hurricanes) and general environments lacking prior geographical information of BSs. Unlike conventional drive testing, a patrol UAV is employed to detect terrestrial BSs without any prior knowledge of their locations or operational status. The real-time detection results are transmitted to nearby surviving BSs or to a remote low-Earth-orbit (LEO) satellite to determine the size and boundaries of the coverage hole.
		    
		    \item UAV scheduling algorithms: Two distinct UAV scheduling schemes are designed according to the size of the detected coverage hole. Specifically, an offline heuristic scheduling algorithm is developed for small and medium-sized networks, whereas an online scheduling algorithm is devised for large-scale networks. To ensure seamless coverage recovery, the lower and upper bounds on the number of ABSs are derived based on circle-covering theory.
		    
		    \item Multi-UAV control with collision avoidance: To enhance the robustness and reliability of swarm operations, a multi-UAV movement-control strategy is developed using a potential function-based approach under a leaderless structure. This strategy guarantees collision-free formation maintenance and provably global asymptotic stability, even without the Lipschitz condition.
		\end{enumerate}

    \subsection{Paper Organization}
        The remainder of this paper is organized as follows. Section~\ref{sec:model} proposes a novel air-to-ground wireless communication system model for coverage hole detection and recovery. Section~\ref{sec: chd} designs algorithms for coverage hole detection and recovery, while Section~\ref{sec:control} devises a movement control strategy with collision avoidance for UAV swarms. Simulation and numerical results are presented and discussed in Section~\ref{sec:simulation}. Finally, Section~\ref{sec:conclusion} concludes the paper.
    
        {\it Notation:} Scalars, vectors, and matrices are denoted by italic, lower-, and uppercase bold letters, respectively. The symbols $\mathbb{R}^{n}$, $\mathbb{R}_{+}$, and $\mathbb{N}_{+}$ indicate the real space of dimension $n$, the set of positive real numbers, and the set of natural numbers, respectively. The symbols $|\cdot|$ and $\|\cdot\|$ denote the absolute value of a number or the cardinality of a set and the $\ell_2$-norm of a vector/matrix, respectively. The superscripts $(\cdot)^{\rm{T}}$ represent the transpose, and the symbol $\mathbb{B}(\bm{x},r)$ refers to an open Eucliden circle with center $\bm{x}$ and radius $r$. The operator $\mathbb{E} [\cdot]$ computes the expectation of a random variable, $\dot{\bm{x}}(t)$ takes the first-order differential of the function $\bm{x}(t)$ with respect to $t$, and $\oplus$ calculates the Minkovski sum of convex polygons. In particular, the functions ${\rm{vol}}(\cdot)$ and ${\rm{sgn}}(\cdot)$ denote the volume of geometric bodies and the signum function, respectively. The Gamma function is defined as $\Gamma(a) \triangleq \int_{0}^{\infty} t^{a - 1} \exp(- t) \,  {\rm d} t$, which can be computed using built-in functions in regular numerical software, such as MATLAB and Mathematica.

\section{System Modeling and Workflow} \label{sec:model}
    We consider downlink transmission in a cellular network deployed over a given area, where coverage holes may arise due to environmental blockages or natural disasters. The terrestrial BSs, each with height $H_{\rm{BS}}$, are modeled as a homogeneous Poisson point process (PPP) $\Phi_{B}$ with density $\lambda_{B}$. Mathematically, a coverage hole $A$ is defined as the geographical region where the received signal-to-interference-plus-noise ratio (SINR) falls below a predefined threshold $\gamma_{\rm{th}}$, i.e., 
	\begin{equation} \label{Eq:CoverageHole}
		A = \{(x, y) \in \mathbb{R}^{2} \mid {\rm{SINR}}(x, y) < \gamma_{\rm{th}}\}.
	\end{equation}

    Each ABS is equipped with a single antenna and operates at an altitude within the range $[H_{1}, H_{2}]$, where $H_{2} > H_{1} \geq H_{\rm{BS}}$. In contrast to the PPP-distributed terrestrial BSs, the initial locations of ABSs follow an aggregated spatial pattern, modeled by a cluster process such as the Neyman--Scott process\cite[Ch.~5]{Chiu2013}. Once scheduled, the ABSs depart from a designated gathering point and fly toward the target coverage-hole region.

\begin{figure}[!t]
	    \centering
	    \includegraphics[width=0.95\linewidth]{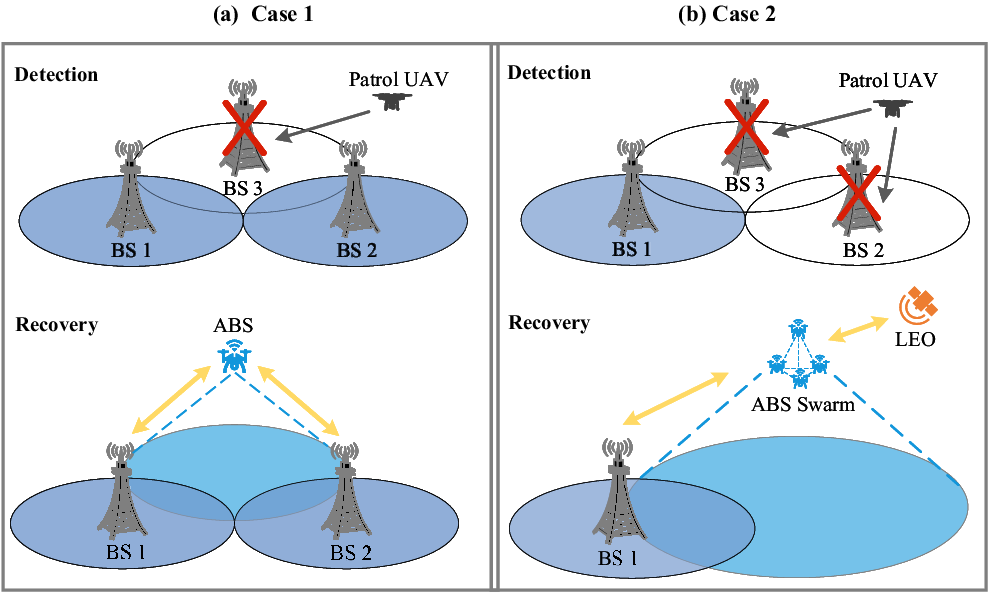}
	    \caption{Illustration of coverage-hole detection and recovery: (a) a small coverage hole is detected by the patrol UAV and recovered by a single ABS with wireless backhaul provided by nearby terrestrial BSs~1 and~2; and (b) a large coverage hole is detected by the patrol UAV and recovered by an ABS swarm with wireless backhaul provided by terrestrial BS~1 and a remote LEO satellite.}
	    \label{Fig-1}
    \end{figure}

    For illustration, Fig.~\ref{Fig-1} depicts two representative scenarios for terrestrial coverage-hole detection and recovery. The left panel shows a small coverage hole detected by the patrol UAV and subsequently recovered by a single ABS with wireless backhaul provided by nearby terrestrial BSs~1 and~2. In contrast, the right panel illustrates a larger coverage hole detected by the patrol UAV and recovered by an ABS swarm, where wireless backhaul is established via terrestrial BS~1 and a remote LEO satellite. To achieve both cost efficiency and recovery effectiveness, the proposed framework leverages UAV mobility to adaptively deploy either a single ABS (as in the left panel of Fig.~\ref{Fig-1}) or an ABS swarm (as in the right panel), depending on the estimated size of the detected coverage hole.

    \subsection{Channel Model}
        We now discuss the transmission models in the detection and recovery stages. 

        In the \emph{detection stage}, the patrol UAV flies at a relatively low altitude to approximate the channel conditions experienced by terrestrial users. It receives signals transmitted by terrestrial BSs and calculates the received SINR. The communication links $\bm{h}_{\rm{C2A}}$ from terrestrial BSs to the patrol UAV may experience various small-scale fading effects. Hence, we adopt the Nakagami-$m$ model to capture a wide range of fading environments. Accordingly, the small-scale fading power gain between a terrestrial BS and the patrol UAV follows the Gamma distribution with PDF
		\begin{equation} \label{Eq: nakagami}
			f_{\|\bm{h}_{\rm{C2A}}\|}(x) = \frac{2 m^{m} x^{2 m - 1}} {\Gamma(m) \Omega^{m}} \exp \!\left(-\frac{m}{\Omega} x^{2}\right), \quad m \geq 0.5,~ x \geq 0,
		\end{equation}
		where the subscript ``$\rm C2A$'' of $\bm{h}_{\rm C2A}$ denotes the link from a terrestrial cellular BS to a patrol UAV, and $m$ and $\Omega$ represent the shape factor and average power, respectively. 
        
        	In the \emph{recovery stage}, by contrast, the small-scale fading between aerial ABSs and ground UEs is negligible due to the dominant LoS propagation, which is consistent with measurement-based studies of air-to-ground channel models for low-altitude platforms reporting a high LoS probability and relatively weak small-scale fluctuations in the considered height and environment ranges~\cite{Zheng2024Aerial, Khawaja2019Survey, Rodriguez-Pineiro2021A2G}.
        
    \subsection{System Workflow}
        The overall workflow of the proposed system is summarized as follows. A patrol UAV routinely cruises along a predefined trajectory to detect potential coverage holes. Upon completing detection, it sends the results to a remote LEO satellite or a nearby BS, which determines whether to dispatch a single ABS or an ABS swarm for coverage recovery, depending on the hole size. 
        
        \begin{figure}[!t]
        	\centering
        	\includegraphics[width=1.0\linewidth]{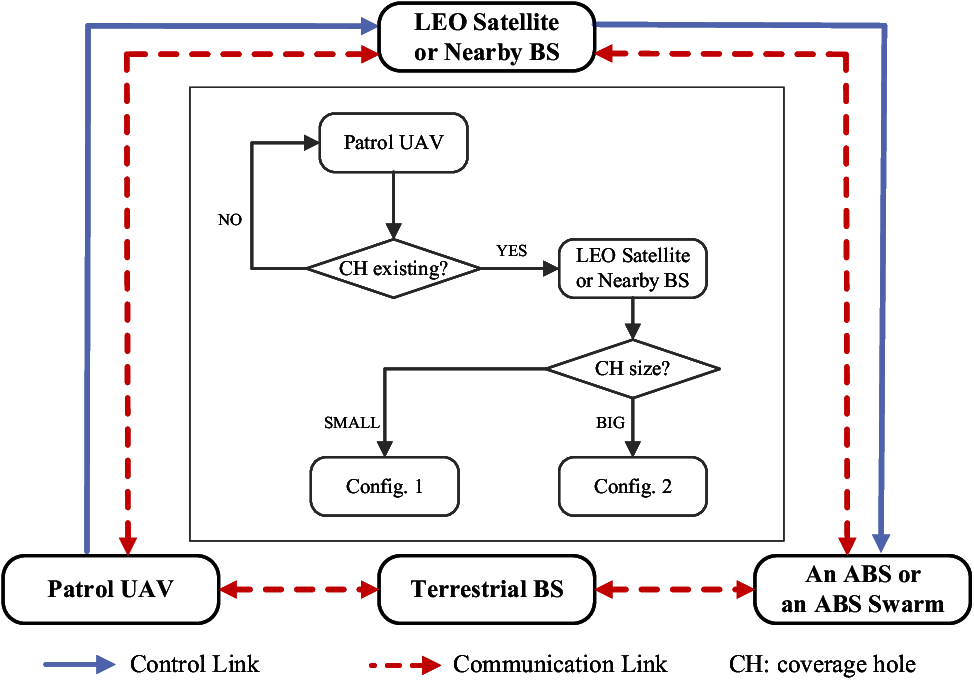}
        	\caption{System workflow under study, where blue links indicate control information flow and red links represent communication data flow. Config.~1 corresponds to a single ABS serving a small coverage hole, while Config.~2 involves an ABS swarm serving a large one.}
        	\label{Fig-2}
        \end{figure}

        For illustration, Fig.~\ref{Fig-2} shows the system workflow. From the communications perspective, the patrol UAV continuously receives signals from terrestrial BSs, computes the received SINR, and compares it with the threshold $\gamma_{\rm th}$ to identify coverage holes. In practice, the patrol UAV hovers at predefined checkpoints to perform reliable detection. If a coverage hole is identified, the geographic information of the corresponding checkpoint is sent to a nearby BS or an LEO satellite, which then schedules the appropriate ABS deployment.

		From the control perspective, the patrol UAV, the LEO satellite or nearby BS, and the ABSs form a closed control loop. Upon receiving detection information from the patrol UAV, the controller (LEO or BS) analyzes the detection results at two consecutive checkpoints surrounding the hole and decides whether to deploy a single ABS or an ABS swarm. The scheduled ABS(s) then fly to the designated checkpoint to recover the coverage hole. 
		
		Although the above description assumes a single patrol UAV, the proposed framework can be readily extended to multiple patrol UAVs operating cooperatively, as the detection checkpoints are supposed to be spatiotemporally independent.

\section{Coverage Hole Detection and Recovery} \label{sec: chd}
     This section first discusses the principle of coverage-hole detection by a patrol UAV and then the principle of coverage-hole recovery by a single ABS or an ABS swarm, effectively recovering potential coverage holes in large-scale wireless networks. Subsequently, an offline heuristic method and an online scheduling algorithm are developed to support integrated coverage hole detection and recovery. Finally, using the circle-covering theory, we establish lower and upper bounds on the number of ABSs required to achieve seamless coverage in a given area.  
      
    \subsection{The Principle of Coverage Hole Detection} \label{sec: chd-patrol AAV}
        Given a SINR threshold $\gamma_{\rm{th}}$, we can exploit \eqref{Eq:CoverageHole} to determine whether the patrol UAV is in the coverage area or not. Based on the transmission models discussed in Section~\ref{sec:model}, the received SINR of the patrol UAV at a checkpoint can be computed as 
        \begin{align}\label{Eq: SINR-UE}
        	{\rm{SINR}}_{{\rm C2A}, i} 
        	= \frac{P_{{\rm BS}, i}^{\rm{tx}} \|\bm{h}_{{\rm C2A}, i}\|^{2} \, d_{i}^{- \alpha_{\rm B}}} {\sum_{m = 1, m \neq i}^{M} P_{{\rm BS}, m}^{\rm{tx}} \|\bm{h}_{{\rm C2A}, m}\|^{2} \, d_{m}^{- \alpha_{\rm B}} + n},
        \end{align}
        where term in numerator denotes the received signal power from the $i^{\rm{th}}$ serving BS to the patrol UAV; $P_{{\rm BS}, i}^{\rm{tx}}$ refers to the transmit signal power of the $i^{\rm{th}}$ BS, $\bm{h}_{\rm{C2A}, i}$ refers to the small-scale fading, and $d_{i}^{- \alpha_{\rm B}}$ indicates the large-scale path loss from $i^{\rm{th}}$ serving BS to the patrol UAV, with the Euclidean distance $d$ between them and the path-loss exponent $\alpha_{\rm B} > 2$. Also, the term in denominator of \eqref{Eq: SINR-UE} represents the total interference from the neighboring cells; the parameter $P_{{\rm BS}, m}^{\rm{tx}}$ means the transmit power of the $m^{\rm{th}}$ BS, with $1 \le m \le M$; $\bm{h}_{{\rm{C2A}}, m}$ stands for the small-scale fading between the $m^{\rm{th}}$ BS and the patrol UAV with $d_{m}$ being the Euclidean distance between them. Finally, the term $n$ in the denominator of the first equality of \eqref{Eq: SINR-UE} represents the power of circular symmetric additive white Gaussian noise.
        
        In principle, the checkpoints along the trajectory of the patrol UAV are randomly selected according to a homogeneous PPP with intensity $\lambda_{\rm{CP}}$, and then exclude any point whose distance to its nearest neighboring point is less than a given positive constant $D$ \cite[Ch. 3]{Matern1986}, which is determined by the coverage radius of a single UAV. In other words, the location of checkpoints $\Phi_{\rm{CP}}$ essentially follows the Mat\'{e}rn hard-core process with $\mu_{\rm{CP}} = \lambda_{\rm{CP}} \exp\left(- 2 \pi \lambda_{\rm{CP}} D^{2}\right)$. This setting enables the detection of potential coverage holes, as PPP can achieve a uniform distribution of checkpoints. At the same time, the thinning transformation ensures that the final checkpoints are distributed with minimum distance $D$, thereby reducing the number of checkpoints and system cost. Considering the maneuverable energy consumption of the UAV and the efficiency of coverage hole detection, the patrol UAV should traverse all checkpoints without repeating according to the nearest distance criterion. Therefore, for the patrol UAV's path planning, many existing algorithms for solving Hamiltonian path problems can be utilized to meet the above requirements, such as the greedy algorithm based on the nearest-neighbor criterion~\cite{Kuehn2012Survey}.
        
        \begin{figure}[!t]  
        	\centering  
        	\includegraphics[width=0.75\linewidth]{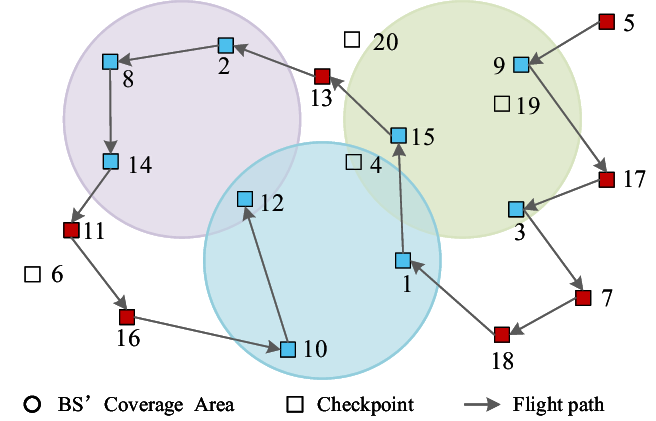}   
        	\caption{The trajectory of the patrol UAV for coverage hole detection. The blue/red squares denote the preset checkpoints of the patrol UAV, and the arcs show the path determined by the greedy algorithm based on the nearest-neighbor criterion. The color of each checkpoint indicates its coverage status: blue denotes being within the coverage area, whereas red denotes being outside it.}  
        	\label{Fig-3}  
        \end{figure}  
        
        For illustrative purposes, Fig.~\ref{Fig-3}  shows a trajectory of the patrol UAV for coverage hole detection. In particular, three circular areas denote the coverage areas of three adjacent terrestrial BSs, and the squares represent the original checkpoints. Moreover, the red/blue squares indicate the $16$ checkpoints of the second selection that belong to $\Phi_{\rm{CP}}$ (i.e., the points $4$, $6$, $19$, and $20$ are excluded), and the arcs refer to the Hamiltonian path for traversing all checkpoints. The patrol UAV flies along the trajectory and puts a label (red or blue) at each preset checkpoint by \eqref{Eq:CoverageHole} and \eqref{Eq: SINR-UE} together with a given $\gamma_{\rm{th}}$. Thus, a directed graph $\mathcal{G}_{\rm{CP}}$ that consists of $16$ points as shown in Fig.~\ref{Fig-3}  is formed. In particular, the checkpoints outside the coverage areas are in red (i.e., the received SINR of the patrol UAV falls below $\gamma_{\rm{th}}$). In contrast, the other checkpoints within the coverage areas are colored blue (i.e., the received SINR of the patrol UAV is or exceeds $\gamma_{\rm{th}}$).

    \subsection{Principle of Coverage Hole Recovery} \label{sec:chd-AAV_swarm}
	    Depending on the size of a detected coverage hole, two ABS deployment configurations are considered:
		\begin{itemize}
		    \item \textbf{Config.~1:} a single ABS; 
		    \item \textbf{Config.~2:} an ABS swarm consisting of four UAVs arranged in a regular tetrahedral formation.
		\end{itemize}
	
	    Config.~1 is applied to recover a \emph{small} coverage hole, characterized by isolated red checkpoints (e.g., checkpoints $5$, $17$, and $13$ in Fig.~\ref{Fig-3}). Config.~2 is used for \emph{large} coverage holes, identified by two or more consecutive red checkpoints (e.g., checkpoints $7$ and $18$, or $11$ and $16$ in Fig.~\ref{Fig-3}). For large-scale recovery, Config.~2 adopts a four-UAV regular tetrahedral formation as the basic cooperative unit, following our previous works \cite{Li2020B, Zhao2022Unified, Fan2023}, which was validated with respect to 3D coverage and motion control. In this configuration, the four UAVs perform CoMP transmission to serve ground UEs.
	
		For a ground UE located within a detected coverage hole, the received SINR from the ABS(s) can be expressed as
		\begin{equation} \label{Eq:SIR-UE}
			{\rm SINR}_{\rm A2G} 
			\triangleq \frac{S_{\rm A2G}}{I_{\rm A2G} + n} 
			\approx \frac{1}{n}P_{\rm ABS}^{\rm tx} \sum_{k = 1}^K \|\bm{h}_{{\rm A2G}, k}\|,
		\end{equation}
		where $\rm{A2G}$ denotes the link from the ABS(s) to the ground UE, $S_{\rm A2G}$ is the received signal power from the ABS(s), $P_{\rm ABS}^{\rm tx}$ is the transmit power of the ABS(s), and $\|\bm{h}_{{\rm A2G}, k}\|$ is the channel power gain between the $k$th ABS and the UE. Here, $K = 1$ for Config.~1 and $K = 4$ for Config.~2. The term $I_{\rm A2G}$ represents interference from neighboring BSs, which is negligible since the UE is located inside a coverage hole and can only be served by the ABS(s). Finally, $n$ denotes the power of circularly symmetric additive white Gaussian noise.
		
		The determination of coverage-hole size is based on the global detection information represented by the directed graph $\mathcal{G}_{\rm CP}$. Holes are categorized as small or large depending on whether the red checkpoints appear separately or consecutively (cf.~Fig.~\ref{Fig-3}). Once the locations and sizes are identified, a satellite or nearby BS schedules the ABS deployment. In Config.~1, the ABS is positioned at the exact location of the red checkpoint. In Config.~2, the centroid of the ABS swarm is aligned with the corresponding red checkpoint. For brevity, `R' and `B' are used to denote red and blue checkpoints, respectively, unless otherwise specified.
		
		\begin{algorithm}[!t]
			\small 
			\setstretch{0.9}
			\caption{An Offline Scheduling Algorithm}
			\label{Alg: CHD-GL}
			\begin{algorithmic}[1]
				\REQUIRE Locations of BSs $\Phi_{B}$ and checkpoints $\Phi_{\rm{CP}}$, channel parameters $\bm{h}_{\rm C2U}$, $\alpha_{\rm{B}}$, $\bm{h}_{\rm A2G}$, $\alpha_{\rm{A, L}}$ and $\alpha_{\rm{A, N}}$ , Tx power $P^{\rm{tx}}_{\rm{BS}}$ and $P^{\rm{tx}}_{\rm{ABS}}$, and SINR threshold $\gamma_{\rm{th}}$;
				\ENSURE Configurations and the locations of ABSs
				\STATE Use existing Hamiltonian path planning algorithm to construct a directed graph $\mathcal{G}_{\rm{CP}}$;
				\STATE Give labels `R' or `B' to the nodes on $\mathcal{G}_{\rm{CP}}$ as per the idea in the last paragraph of Section~\ref{sec: chd-patrol AAV};
				\FOR{the node $i$ with label `R' on $\mathcal{G}_{\rm{CP}}$}
				\IF{the consecutive nodes of node $i$ all with label `B'}
				\STATE The coverage hole around node $i$ is `small';
				\ELSE
				\STATE The coverage hole around node $i$ is `big';
				\ENDIF
				\ENDFOR
				\IF{the coverage hole around node $i$ is `small'}
				\STATE Config.~1 is adopted, and the location of a single ABS is above node $i$;
				\ELSIF{the coverage hole around node $i$ is `big'}
				\STATE Config.~2 is adopted, and the location of the ABS swarm's centroid is above node $i$;
				\ENDIF
			\end{algorithmic}
		\end{algorithm} 

    \subsection{Algorithms for Coverage Hole Detection and Recovery} \label{sec: chd-algorithm}
       \subsubsection{\underline{An Offline Heuristic Algorithm}}
       
       After the patrol UAV traverses its preset trajectory and the directed graph $\mathcal{G}_{\rm{CP}}$ is attained, we can use the simple idea of ``first-discovery-first-recovery'' to recover all detected holes. Accordingly, an offline heuristic method is formalized in {\bf Algorithm~\ref{Alg: CHD-GL}}. It mainly consists of two parts: Lines 1--9 detect the coverage holes and determine their sizes, and Lines 10--14 schedule ABSs to recover them. Due to insensitive scheduling delay, this offline Algorithm~\ref{Alg: CHD-GL} is suitable for {\it small or medium} wireless networks.
    
       \subsubsection{\underline{An Online Scheduling Algorithm}}
          
\begin{figure}[!t]
       	  \centering
       	  \includegraphics[width=1.0\linewidth]{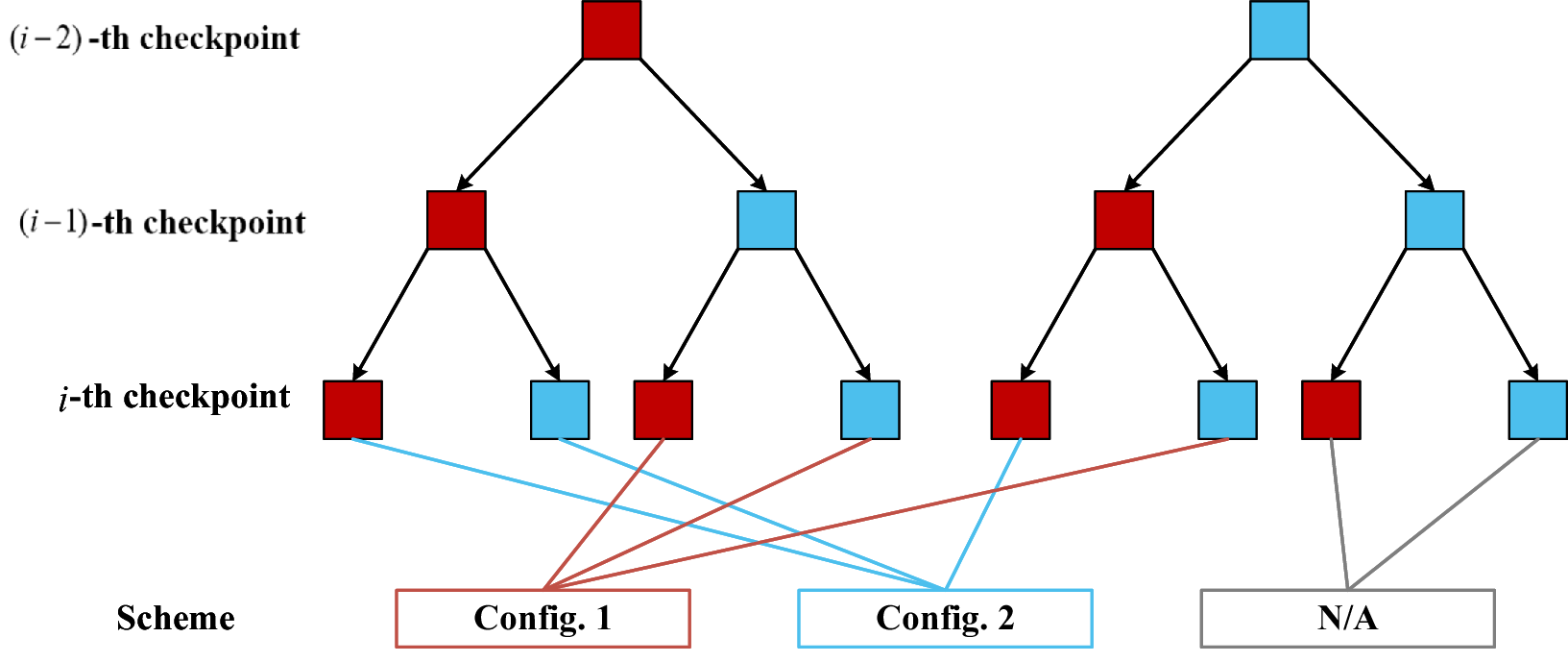}
       	  \caption{An online scheduling policy determined by the detection information of three consecutive checkpoints.}
       	  \label{Fig-4}
       \end{figure}

       For {\it large-scale} wireless networks, the patrol UAV requires a very long time to traverse its preset trajectory, resulting in intolerable service delay for terrestrial users. To address this issue, we have designed an online scheduling algorithm to enhance scheduling efficiency. In principle, by \eqref{Eq:CoverageHole}, the detected coverage holes can be expressed in planar form. Accordingly, the patrol UAV's detection process shown in Fig.~\ref{Fig-3} can be discretized into a sequence of each consisting of three consecutive checkpoints, and the ABS scheduling policy for the first two of each three consecutive checkpoints is determined by their detection results (c.f. Fig.~\ref{Fig-4}). This idea is primarily inspired by triangulation theory in stochastic geometry, which has numerous applications, for example, in global positioning systems, geographical information systems, and robotic localization \cite[Ch. 1]{Hjelle2006}, and recently in wireless communications \cite{Xia2018}.

       Fig.~\ref{Fig-4} exhausts all cases regarding the scheduling policy determined by the detection results of three consecutive checkpoints. The ABS scheduling decision of the first two checkpoints, i.e., the $(i - 2)^{\rm th}$ and $(i - 1)^{\rm th}$ checkpoints, is made based on the information about the coverage hole obtained from every three consecutive checkpoints, i.e., the $(i - 2)^{\rm th}$, $(i - 1)^{\rm th}$ and $i^{\rm th}$ checkpoints. It is evident that Config.~1 is deployed when the color labels of three checkpoints belong to the set \{RBR, RBB, BRB\}, and Config.~2 is deployed with the set \{RRR, RRB, BRR\}. Since the first two checkpoints fully determine this scheduling scheme, no action will be taken if the labels belong to the set \{BBR, BBB\}. As a result, an online scheduling algorithm is formalized in {\bf Algorithm~\ref{Alg:CHD-RT}}. This algorithmic idea could, in principle, be realized via machine learning and other AI-based approaches; we leave these considerations for future work.
       
       \begin{remark}[Choice of checkpoint sequence length]
       	  In Algorithm~2, we adopt sequences of three consecutive checkpoints.  This value is not claimed to be optimal, but it is the minimal length that enables the proposed online decision rule to reliably distinguish small holes from large ones. With only two checkpoints, different spatial patterns (e.g., staggered ``R''/``B'') lead to ambiguous observations. In contrast, more extended sequences may improve robustness at the cost of slower reaction times and the potential masking of short yet critical holes.
       \end{remark}
    
       \begin{algorithm}[!t] 
       	\small
       	\setstretch{0.95}
       	\caption{Online ABS Scheduling Algorithm} 
       	\label{Alg:CHD-RT}
       	\begin{algorithmic}[1]
       		\REQUIRE Locations of terrestrial BSs $\Phi_{B}$ and checkpoints $\Phi_{\rm CP}$; channel parameters $\bm{h}_{\rm C2U}$, $\alpha_{\rm B}$, $\bm{h}_{\rm A2G}$, $\alpha_{\rm A,L}$, $\alpha_{\rm A,N}$; transmit powers $P^{\rm tx}_{\rm BS}$ and $P^{\rm tx}_{\rm ABS}$; SINR threshold $\gamma_{\rm th}$.
       		\ENSURE ABS configurations and deployment locations.
       		
       		\STATE Construct a directed graph $\mathcal{G}_{\rm CP}$ using an existing Hamiltonian path planning algorithm over the checkpoints.
       		
       		\FOR{each checkpoint $i = 3 : \max$}
       		\IF{the color labels of the consecutive three checkpoints belong to $\{ \text{RBR}, \text{RBB}, \text{BRB} \}$}
       		\STATE Deploy \textbf{Config.~1}: a single ABS.
       		\STATE Place the ABS above the first two checkpoints labeled `R'.
       		\ELSIF{the color labels of the consecutive three checkpoints belong to $\{ \text{RRR}, \text{RRB}, \text{BRR} \}$}
       		\STATE Deploy \textbf{Config.~2}: an ABS swarm with tetrahedral formation.
       		\STATE Place the centroid of the ABS swarm above the first two checkpoints labeled `R'.
       		\ENDIF
       		\ENDFOR
       	\end{algorithmic} 
       \end{algorithm}
       
       \subsubsection{\underline{Online vs. Offline Scheduling}}       
       We compare the online Algorithm~\ref{Alg:CHD-RT} with the offline Algorithm~\ref{Alg: CHD-GL}. The offline scheme exploits global information over the whole time horizon and can approach optimal resource allocation, but it performs discovery and recovery sequentially. Consequently, for a point in a coverage hole, the interruption time under Algorithm~\ref{Alg: CHD-GL} is essentially the sum of the discovery and the subsequent recovery periods. In contrast, Algorithm~\ref{Alg:CHD-RT} operates on a local checkpoint sequence and interleaves detection and recovery, enabling recovery to start as soon as red checkpoints appear. This overlap reduces average interruption time while maintaining bounded computational complexity.
       
       To quantify this gain, consider a square region of side length $a$, where all UAVs are located at one corner as the base, and the patrol UAV follows a Hamiltonian path over the checkpoints, as discussed in Section~III-A. By the Beardwood–Halton–Hammersley theorem for the Euclidean space~\cite{Beardwood1959}, for large checkpoint density, the corresponding path length satisfies
       \begin{equation}
       	\mathbb{E}[L_{\rm CP}] \approx \beta\, a^{2}\sqrt{\lambda_{\rm CP}},
       \end{equation}
       where $\beta$ is a universal constant for the Euclidean plane. Hence, the total discovery time scales as
       \begin{equation} \label{Eq: cp-time}
       	\mathbb{E}[T_{\rm CP}] \approx \frac{\beta a^{2}}{v}\, \sqrt{\lambda_{\rm CP}},
       \end{equation}
       where $v$ is the speed of the UAV.
       
       Under Algorithm~1, all checkpoints are first discovered (taking time $T_{\rm CP}$), and ABSs are dispatched afterwards. For a red checkpoint at distance $r$ from the base, the completion time is
       \begin{equation}
       	  T_{\rm Alg1}(r) = T_{\rm CP} + \frac{r}{v}.
       \end{equation}
       Averaging over $R$ (the distance from the base to a randomly selected checkpoint), and using the scaling of $\mathbb{E}[T_{\rm CP}]$ from~\eqref{Eq: cp-time} yields
       \begin{equation} \label{Eq: time-of-alg1}
       	  \mathbb{E}[T_{\rm Alg1}] = \frac{\beta a^{2}}{v} \sqrt{\lambda_{\rm CP}} + \frac{\mathbb{E}[R]}{v}.
       \end{equation}
       Under Algorithm~2, as soon as a checkpoint is discovered and identified as red, the ABS is dispatched immediately. Along the Hamiltonian path with a nearest-distance rule, the discovery time of a checkpoint at distance $r$ can be approximated as  $t(r) \approx T_{\rm{CP}} F_R(r)$, where $F_R$ is the cumulative distribution function of $R$. Thus, the corresponding completion time is  
       \begin{eqnarray}
       	T_{\rm Alg2}(r) \approx T_{\rm{CP}} F_R(r) + \frac{r}{v}.
       \end{eqnarray}
       Taking expectation over $R$ yields
       \begin{equation}
       	\mathbb{E}[T_{\rm Alg2}]
       	\approx \mathbb{E}[T_{\rm CP}]\,\mathbb{E}[F_R(R)] + \frac{\mathbb{E}[R]}{v}.
       \end{equation}
       By the probability integral transform~\cite[Chap.~3]{Billingsley1995}, since $R$ is a continuous random variable with CDF $F_R$, the random variable $F_R(R)$ is uniformly distributed over $[0,1]$. Hence,
       \begin{equation} \label{Eq: time-of-alg2}
       	  \mathbb{E}[T_{\rm Alg2}] \approx \frac{1}{2}\frac{\beta}{v} a^{2}\sqrt{\lambda_{\rm{CP}}} + \frac{\mathbb{E}[R]}{v}.
       \end{equation}
   
       Compared with~\eqref{Eq: time-of-alg1} and \eqref{Eq: time-of-alg2}, we can get the Algorithm~2 reduces the average completion time of red checkpoints by approximately half of the total discovery time, and this gain scales as \(a^{2}\sqrt{\lambda_{\rm{CP}}}/v\), becoming more significant as the checkpoint density $\lambda_{\rm{CP}}$ and the size of the region length $a$ increases. In summary, the online scheduling strategy offers a strictly better delay–performance tradeoff than the offline benchmark in large-scale, densely monitored scenarios.

    \subsection{Minimum Recommended Number of ABSs} \label{sec:num-aBS}
\begin{figure}[!t]
			\centering
			\captionsetup[subfigure]{margin = 5pt} 
			\subfloat[Only terrestrial BSs.]{
				\label{Fig-5a}
				\includegraphics[width=0.35\linewidth]{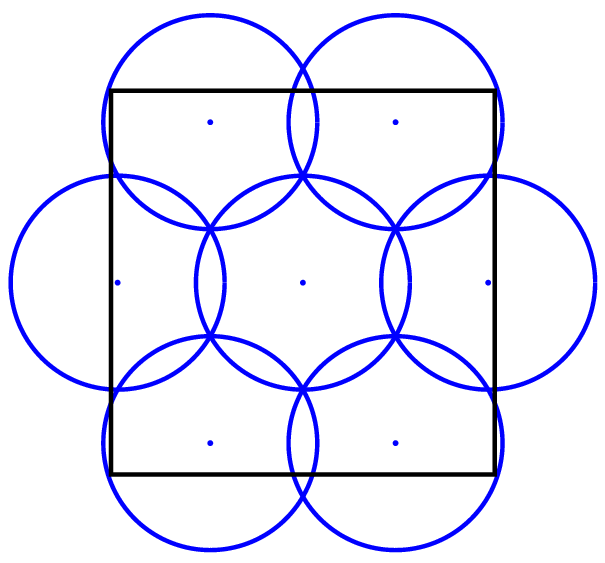}
			}
			\subfloat[The terrestrial BSs and a single ABS of Config.~1.]{
				\label{Fig-5b}
				\includegraphics[width=0.35\linewidth]{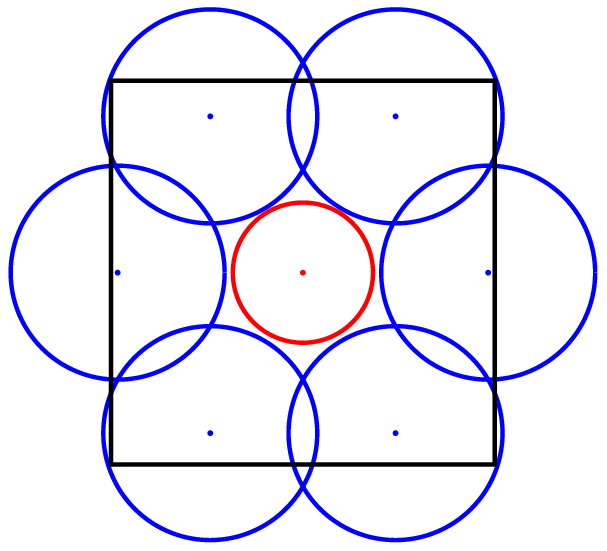}
			}  \\ 
			\subfloat[The terrestrial BSs and an ABS swarm of Config.~2.]{
				\label{Fig-5c}
				\includegraphics[width=0.35\linewidth]{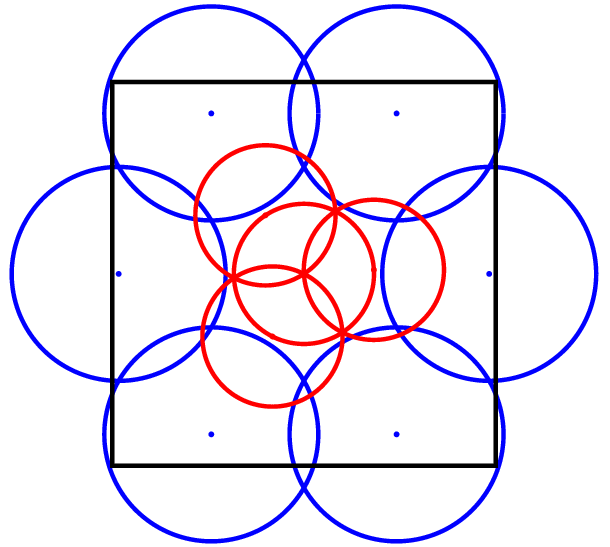}
			} 
			\subfloat[Only ABSs of Config.~1.]{
				\label{Fig-5d}
				\includegraphics[width=0.35\linewidth]{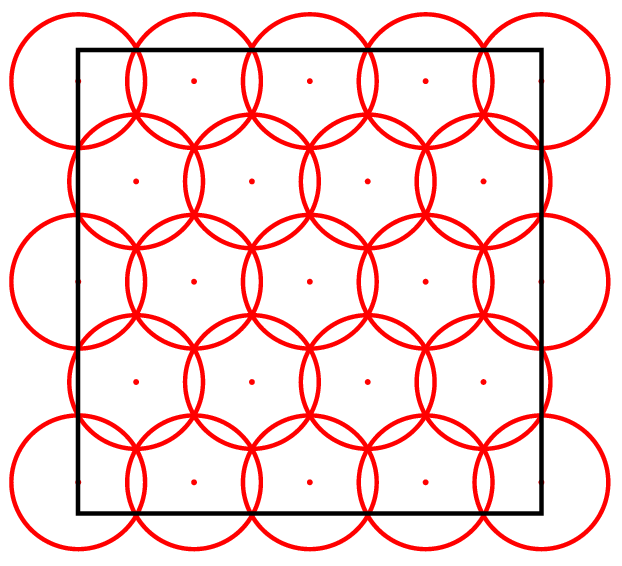}
			}
			\caption{A sketch of the circle covering problem, where the blue and red circles refer to the coverage area of terrestrial BSs and ABSs, respectively.}
			\label{Fig-5}
		\end{figure}

		To implement the proposed recovery method, we must determine the number of ABSs required to achieve seamless coverage over a given area. For mathematical tractability, the BS deployment is formulated as a \emph{circle covering problem}, where the entire network coverage area is filled by multiple circles of equal radius, as illustrated in Fig.~\ref{Fig-5a}. Our objective is to estimate the number of ABSs required to restore the desired quality of service, as shown in Fig.~\ref{Fig-5b} and Fig.~\ref{Fig-5c}. Since the exact size of the coverage hole remains unknown until the patrol UAV completes detection, sufficient ABSs should be prepared for the worst case, as illustrated in Fig.~\ref{Fig-5d}.
		
		\begin{theorem}[Bounds on the Number of BSs Required for Seamless Coverage] \label{Thm:num-aBS}
			Let $\mathcal{V} \in \mathbb{R}^{2}$ denote the coverage region that can be fully covered by $N_c(R)$ BSs, each with an effective coverage radius $R$. Then,
			\begin{equation} \label{eq:fully-cover}
				\underline{N_c} \triangleq 
				\frac{{\rm vol}(\mathcal{V})}{{\rm vol}(\mathcal{B}(o, R))} 
				\leq 
				N_c(R) 
				\leq 
				\frac{{\rm vol}(\mathcal{V} \oplus \mathcal{B}(o, R/2))}{{\rm vol}(\mathcal{B}(o, R/2))} 
				\triangleq 
				\overline{N_c}.
			\end{equation}
		\end{theorem}
		
		\begin{IEEEproof}
		    See Appendix~\ref{Appendix:num}.
		\end{IEEEproof}
		
		Notably, $R$ is a key parameter in Theorem~\ref{Thm:num-aBS} and can be determined as follows. Recalling \eqref{Eq:SIR-UE} with a predetermined SINR threshold $\gamma_{\rm th}$ and assuming the UE is located at the origin, the coverage radius of an ABS is given by
		\begin{equation}
			\|\bm{r}\|^{2} = 
			\left(\frac{P_{\rm {ABS}}^{\rm tx}}{\gamma_{\rm th} n}\right)^{\!\frac{2}{\alpha_{\rm U}}} - h^{2},
		\end{equation}
		where $d = \sqrt{\|\bm{r}\|^{2} + h^{2}}$ is the 3D Euclidean distance between the ABS and the UE, and $h$ is the ABS altitude. For an ABS swarm, the effective coverage radius depends on both $\|\bm{r}_{i}\|$ and the swarm’s geometric configuration, as illustrated in Fig.~\ref{Fig-5c}. Hence, given $P_{\rm {ABS}}^{\rm tx}$, $\gamma_{\rm th}$, and $h \in [H_{1}, H_{2}]$, the required number of aerial BSs $N_c(R)$ can be approximated.
		
		For practical deployment, we take the expectation of $\|\bm{r}\|$ with respect to the ABS altitude distribution and define the effective coverage radius as
		\begin{equation}\label{Eq:num-aBs}
			R \triangleq \mathbb{E}[\|\bm{r}\|]
			\approx 
			\left[
			\left(\frac{P_{\rm {ABS}}^{\rm tx}}{\gamma_{\rm th}}\right)^{\!\frac{2}{\alpha_{\rm U}}}
			- \mathbb{E}[h_i^{2}]
			\right]^{1/2}.
		\end{equation}
		
		Specifically, for a single ABS (Config.~1), \eqref{Eq:num-aBs} reduces to
		\begin{equation}
			R = \left[
			\left(\frac{P_{\rm {ABS}}^{\rm tx}}{\gamma_{\rm th}}\right)^{\!\frac{2}{\alpha_{\rm U}}}
			- h^{2} \right]^{1/2} 
			\triangleq R_{1}.
		\end{equation}
		For an ABS swarm (Config.~2), if one ABS flies at altitude $h_{a}$ and the other three at $h_{b}$, where $h_{a}, h_{b} \in [H_{1}, H_{2}]$ and $h_{a} > h_{b}$, the effective coverage radius can be approximated~as
		\begin{equation}
			R = R_{1} + \frac{\sqrt{2}}{2}(h_{a} - h_{b}) \triangleq R_{2}.
		\end{equation}
		As the vertical height difference between ABSs is limited, the gap between $R_{1}$ and $R_{2}$ is small. Thus, the number of aerial stations required for seamless coverage can be bounded by
		\begin{align} \label{Eq:num-aBs-new}
			\min \!\left\{ \underline{N_c}(R_{1}), \, 4\underline{N_c}(R_{2}) \right\}
			& \leq 
			N_c(R_{1}, R_{2}) \nonumber\\
			& \leq 
			\max \!\left\{ \overline{N_c}(R_{1}), \, 4\overline{N_c}(R_{2}) \right\}.
		\end{align}
		
		This theoretical result will be validated and compared through simulations in Section~\ref{Section-VA}.


\section{Multi-ABS Control with Collision Avoidance} \label{sec:control}
    To recover the detected coverage holes, we schedule multiple ABSs simultaneously to reach the target spatial positions and, if necessary, form the geometric pattern required by an ABSs swarm. In this case, multi-ABS motion control with collision avoidance must be carefully designed for real-world applications. Accordingly, in this section, we exploit graph theory to model the ABS's communication relationship and nonlinear multi-agent system theory to address the ABS swarm control problem with collision avoidance. Next, we start with the ABS dynamics.

    \subsection{The ABS Dynamics} \label{sec: control-model}
        Suppose there exist $M$ ABSs of Config.~1 and $N$ ABS swarms of Config.~2 at a given moment for coverage recovery, and let $\mathcal{P}$ and $\mathcal{Q}$ denote their respective ABS sets. It is apparent that there are $M + 4N$ ABSs in total because of $|\mathcal{P}| = M$ and $|\mathcal{Q}| = 4N$. Considering the $i^{\rm th}$ ABS with a 3D state vector, $i = 1, \cdots, M + 4N$, its dynamics can be described by  
        \begin{subequations} 
        	\label{Eq: CA-general}
        	\begin{align}
        		\dot{\bm{x}}_{i}(t) & = \bm{v}_{i}(t), \label{Eq: CA-general-a} \\
        		\dot{\bm{v}}_{i}(t) & = {\bm f}_{R} \left(\bm{v}_{i}(t)\right) + {\bm u}_{i}(\bm{x}_{i}(t), \bm{v}_{i}(t)), \label{Eq: CA-general-b}
        	\end{align} 
        \end{subequations}
        where the vector $\bm{x}_{i}(t) \in \mathbb{R}^{3}$ in \eqref{Eq: CA-general-a} denotes the geographic position of the $i^{\rm th}$ ABS, and the first-order differential $\dot{\bm{x}}_{i} (t)$ with respect to $t$ means its velocity given by $\bm{v}_{i} (t)$ in \eqref{Eq: CA-general-a}; the function ${\bm f}_{R} \left(\bm{v}_{i} (t)\right) \in \mathbb{R}^{3}$ in \eqref{Eq: CA-general-b} represents the intrinsic nonlinear dynamics of the $i^{\rm th}$ ABS, and ${\bm u}_{i} (t) \in \mathbb{R}^{3}$ in \eqref{Eq: CA-general-b} denote the control vector of the $i^{\rm th}$ ABS at time $t$, with two parameters $\bm{x}_{i}(t)$ and $\bm{v}_{i}(t)$. For the sake of notational simplicity, all independent variables in \eqref{Eq: CA-general-a}-\eqref{Eq: CA-general-b} are omitted if no confusion arises. 
        
        The intrinsic nonlinear dynamics of each ABS considered in this paper comprise air resistance, gravity, and a portion of the lift that balances gravity. The remainder of the lift is used to maintain the ABS in the vertical direction, as accounted for in the controller design in the next subsection. Mathematically, the function ${\bm f}_{R} \left(\bm{v}_{i}(t)\right)$ in \eqref{Eq: CA-general-b} can be explicitly approximated by \cite[Eq. (2.2)]{Taylor2005} as 
        \begin{equation}\label{Eq: CA-general-fr}
        	{\bm f}_{R} \left(\bm{v}_{i}\right) = - \left(k_{1} \|\bm{v}_{i}\|_{2} + k_{2} \|\bm{v}_{i}\|_{2}^{2}\right) \frac{\bm{v}_{i}}{M_{i} \|\bm{v}_{i}\|}, 
        \end{equation}
        where $M_{i}$ denotes the mass of the $i^{\rm th}$ ABS, and the positive coefficients $k_{1}$ and $k_{2}$ are in the unit of kilograms per second (\si{kg/s}) and kilograms per meter (\si{kg/m}), respectively.

    \subsection{ABS Swarm Control with Collision Avoidance} \label{sec: control-scheme}
\begin{figure}
        	\centering
        	\includegraphics[width=0.8\linewidth]{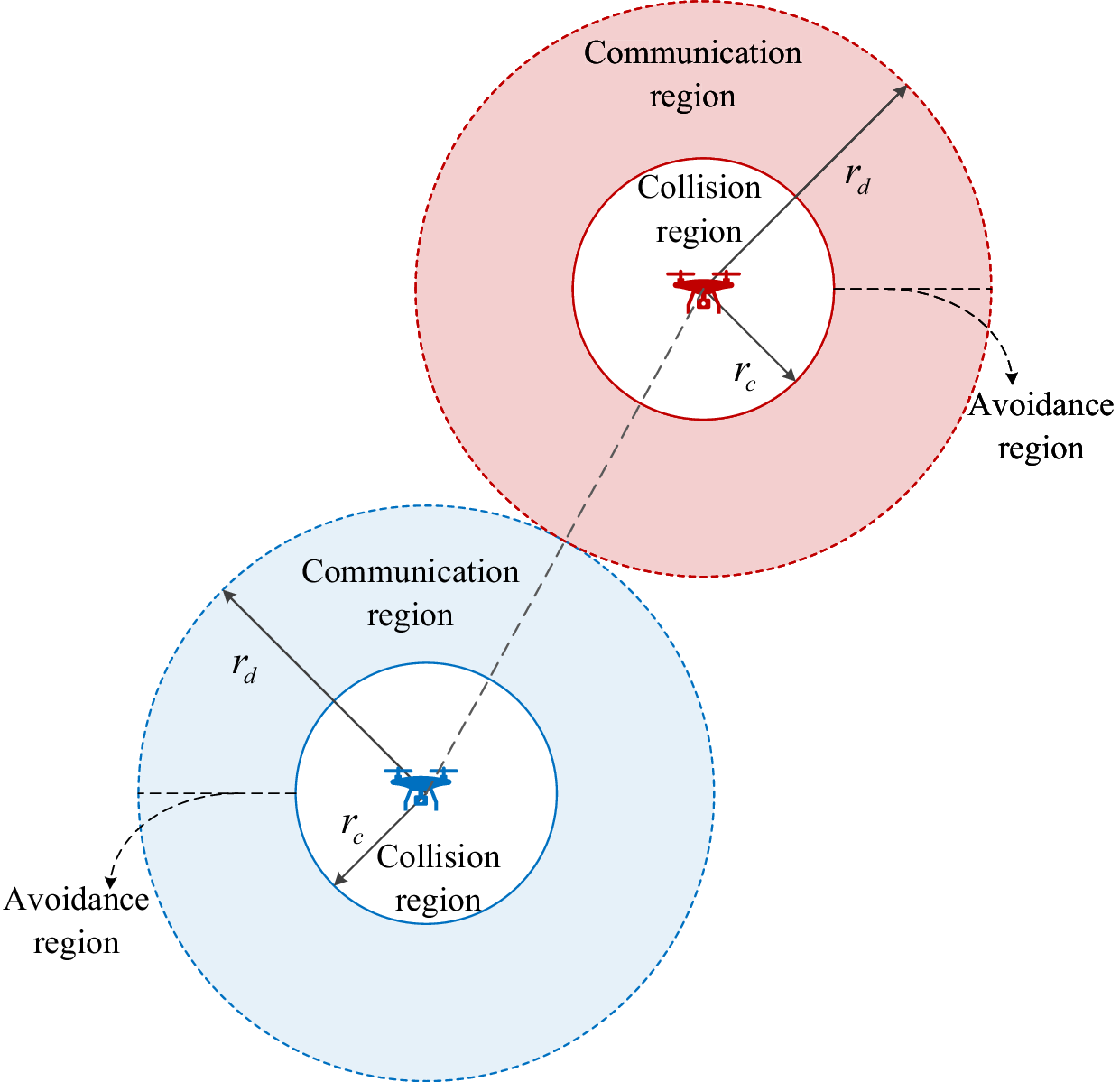}
        	\caption{Illustration of the collision and communication regions for a target ABS and its adjacent one.}
        	\label{Fig-6}
        \end{figure}

        We begin to define the collision and communication regions in the collision-avoidance scenario. As shown in Fig.~\ref{Fig-6}, for the $i^{\rm th}$ ABS located at $\bm{x}_{i}$, the collision region is defined as $\Psi_{i} \triangleq \{\bm{p}_{i} \mid \|\bm{x}_{i} - \bm{p}_{i}\| \leq r_{c}\}$ and the communication region is defined as $\Omega_{i} \triangleq \{\bm{p}_{i} \mid \|\bm{x}_{i} - \bm{p}_{i}\| \leq r_{d}\}$, with $r_{c} \leq r_{d}$. Thus, the mutual avoidance region for the $i^{\rm th}$ ABS can be defined as $\Psi_{i} \cap \Omega_{i}$. If the $j^{\rm th}$ ABS moves into the mutual avoidance region of $i^{\rm th}$ ABS, i.e., $\Omega_{i} \cap \Psi_{j} \ne \varnothing$, for all $i \ne j$, their collision avoidance mechanism will be activated.
        
        \underline{Case 1: The ABS Set of Config.~1.} Consider the $i^{\rm th}$ ABS in the ABS set $\mathcal{P}$, we adopt a control strategy ${\bm u}_{i} \triangleq {\bm u}_{i}^{(\mathcal{P})}$ to solve~\eqref{Eq: CA-general}. According to \cite{Fan2023}, ${\bm u}_{i}^{(\mathcal{P})}$ can be given by
        \begin{equation} \label{Eq: CA-Pu}
        	{\bm u}_{i}^{(\mathcal{P})} = - b_{i} \bm{v}_{i} - \sum_{q \in \mathcal{P} \cup \mathcal{Q}} \frac{\partial V^{ik}\left(\bm{x}_{i}, \bm{x}_{k}\right)}{\partial \bm{x}_{i}} + \bm{u}^{\rm{(adap)}}_{i},
        \end{equation}
        where $b_{i} \geq 0$ denotes the local feedback gain obtained at the $i^{\rm th}$ ABS; the collision avoidance potential function is determined as \cite{Hokayem2010Coordination}, i.e.,
        \begin{equation} \label{Eq: CA-function-V}
        	V^{(i,k)}\left(\bm{x}_{i}, \bm{x}_{k}\right) =
        	\begin{cases}
        		\left(\dfrac{r_{d}^{2} - \|\bm{x}_{i} - \bm{x}_{k}\|^{2}}{\|{\bm{x}_{i}} - \bm{x}_{k}\|^{2} - r_{c}^{2}}\right)^{2}, \\
        		\qquad \ \text{if  } r_{c} \leq \|{\bm{x}_{i}} - \bm{x}_{k}\|^{2} \leq r_{d}; \\
        		0, \quad \ \text{if  } r_{d} \leq \|{\bm{x}_{i}} - \bm{x}_{k}\|^{2},
        	\end{cases}
        \end{equation}
        and the partial derivative of $V^{(i,k)}\left(\bm{x}_{i}, \bm{x}_{k}\right)$ with respect to the each element of $\bm{x}_{i}$, say, $x_{i, \ell} $ with $\ell \in \{1, 2, 3\}$, is given by
        \begin{equation} \hspace{-1em}
        	\frac{\partial V^{(i,k)}\left(\bm{x}_{i}, \bm{x}_{k}\right)}{\partial x_{i, \ell}} = 
        	\begin{cases}
        		\frac{4 \left(r_{d}^{2} - r_{c}^{2}\right) \left(r_{d}^{2} - \|\bm{x}_{i} - \bm{x}_{k}\|^{2}\right) \left(x_{k, \ell} - x_{i, \ell}\right)}{\left(\|{\bm{x}_{i}} - \bm{x}_{k}\|^{2} - r_{c}^{2}\right)^{3}}, \\
        		\qquad \quad \ \text{if  } r_{c} \leq \|{\bm{x}_{i}} - \bm{x}_{k}\|^{2} \leq r_{d}; \\
        		0, \qquad \ \text{if  } r_{d} \leq \|{\bm{x}_{i}} - \bm{x}_{k}\|^{2}. 
        	\end{cases}
        \end{equation}
        The last term of \eqref{Eq: CA-Pu} is $\bm{u}^{{\rm{(adap)}}}_{i} \in \mathbb{R}^{3}$, whose element $u^{\rm{(adap)}}_{i, \ell}$ with $\ell \in \{1, 2, 3\}$, is expressed as 
        \begin{equation} \label{Eq: CA-u3}
        	u^{\rm{(adap)}}_{i, \ell} = c_{i} \cdot
        	\begin{cases}
        		\varepsilon_{i, \ell} \, {\rm sgn} \left(x_{i, \ell} - x_{i, \ell}^{\rm{dest}}\right), \\
        		\qquad \qquad \quad \ \, \text{if  } |x_{i, \ell} - x_{i, \ell}^{\rm{dest}}| > \varepsilon_{i, \ell}; \\
        		x_{i, \ell} - x_{i, \ell}^{\rm{dest}}, \ \text{if  } |x_{i, \ell} - x_{i, \ell}^{\rm{dest}}| \leq \varepsilon_{i, \ell},
        	\end{cases}
        \end{equation}
        where $c_{i}$ and $\varepsilon_{i, \ell}$ are both positive control constants, and $\bm{x}_{i}^{\rm{dest}} \triangleq \left[x_{i, 1}^{\rm{dest}}, x_{i, 2}^{\rm{dest}}, x_{i, 3}^{\rm{dest}}\right]^{\rm{T}}$ denotes the destination position of the $i^{\rm th}$ ABS; and the variables mentioned above have similar meanings to those in~\eqref{Eq: CA-general}, unless specified otherwise. 
        
        Intuitively, the potential function given by \eqref{Eq: CA-function-V} combines an attractive term that pulls each UAV toward its desired formation position and repulsive terms that push it away from nearby UAVs/obstacles, so that the negative gradient term in \eqref{Eq: CA-Pu} can be viewed as a virtual force field guiding the motion.
        
        \underline{Case 2: The ABS Swarm of Config.~2.} 
        The ABSs within the same swarm interact with one another. Thus, the ABS set $\mathcal{Q}$ can be decomposed into $\mathcal{Q} = \left\{\mathcal{Q}_{1}, \cdots, \mathcal{Q}_{N}\right\}$, and each subset $\mathcal{Q}_{n}$ consists of $4$ ABSs, and $\mathcal{Q}_{m} \cap \mathcal{Q}_{n} = \varnothing$, for all $m \ne n$ and $m, n \in \{1, \cdots, N\}$. To model the interaction among ABSs in a swarm, graph theory is employed, and several concepts are introduced before proceeding. More specifically, we model the ABS swarm $\mathcal{Q}_{n}$ as a nonlinear multi-agent system of $4$ agents. The interaction among the agents can be described as a weighted digraph $\mathcal{G} = \{\mathcal{V}, \mathcal{E}\}$, where $\mathcal{V} = \{1, 2, 3, 4\}$ is a set of nodes, and $\mathcal{E} \subseteq \mathcal{V} \times \mathcal{V}$ represents a set of edges. A directed edge $a_{ij}$ in $\mathcal{E}$ is denoted by the ordered pair of nodes $\left(i, j\right)$, which means node $j$ can get information from node $i$, yet not vice versa. Define the weighted adjacency matrix ${\bm A} = \left[a_{i j}\right] \in \mathbb{R}^{4 \times 4}$ with $a_{i j} = 0$ if $i = j$ and $a_{i j} \geq 0$ if $i \neq j$.
        
        Consider the $i^{\rm th}$ ABS in the swarm $\mathcal{Q}_{n} \subset \mathcal{Q}$, we adopt a control strategy ${\bm u}_{i} \triangleq {\bm u}_{i}^{(\mathcal{Q})}$ to solve~\eqref{Eq: CA-general} for this ABS swarm, given by \cite{Fan2023} as 
        \begin{align} \label{Eq: CA-Qu}
        	{\bm u}_{i}^{(\mathcal{Q})} & = - \left(\sum_{j \in \mathcal{Q}_{n}} a_{i j}(\bm{x}_{i}- \bm{x}_{j} - \bm{x}_{i}^{\ast} + \bm{x}_{j}^{\ast}) + b_{i} \bm{v}_{i} \right. \nonumber \\
        	& \quad \left. + \sum_{k \in \mathcal{P} \cup \mathcal{Q}} \frac{\partial V^{(i,k)}\left(\bm{x}_{i}, \bm{x}_{k}\right)}{\partial \bm{x}_{i}} + \bm{u}^{\rm{(adap)}}_{i}\right), 
        \end{align} 
        where $\bm{x}_{i}^{\ast}$ denotes the expected respective relative position of the $i^{\rm th}$ ABS in the swarm formation, and the other variables have similar meanings to those in~\eqref{Eq: CA-general} and \eqref{Eq: CA-Pu}. Compared with \eqref{Eq: CA-Pu}, the first term in the parentheses of \eqref{Eq: CA-Qu} is added to ensure the geometric formation of an ABS swarm.
                
        For notation simplicity, define $\tilde{\bm{x}}_{i} \triangleq \bm{x}_{i} - \bm{x}_{i}^{\rm{dest}}$ as the related position vector between the $i^{\rm th}$ ABS and its destination, and $\tilde{\bm{v}}_{i} \triangleq \dot{\tilde{\bm{x}}}_{i} = \bm{v}_{i}$. It is noteworthy that the two-dimensional (2D) projection of the $i^{\rm th}$ ABS location (i.e., $\bm{x}_{i}^{\rm{dest}}$) is the same as its corresponding checkpoint in set $\mathcal{P}$. As for the ABS swarm in set $\mathcal{Q}$, $\bm{x}_{i}^{\rm{dest}}$ of the $i^{\rm th}$ ABS is not only related to its corresponding checkpoint but also to its respective relative position in the swarm formation (i.e., $\bm{x}_{i}^{\ast}$). Next, combining~\eqref{Eq: CA-general}--\eqref{Eq: CA-Pu} and \eqref{Eq: CA-Qu} and performing some algebraic manipulations, we attain
        \begin{subequations} \label{Eq: CA-system}
        	\begin{align} 
        		\dot{\tilde{\bm{x}}}_{i} & = \tilde{\bm{v}}_{i}, 
        		\label{Eq: CA-general-ra} \\ 
        		\dot{\tilde{\bm{v}}}_{i} & = {\bm f}_{R} \left(\tilde{\bm{v}}_{i}\right) - b_{i} \tilde{\bm{v}}_{i} - \sum_{q \in \mathcal{P} \cup \mathcal{Q}} \frac{\partial V^{(i,k)}\left({\bm{x}}_{i}, {\bm{x}}_{k}\right)}{\partial \bm{x}_{i}}   \nonumber \\
        		& \quad {} - \tilde{\bm{u}}^{\rm{(adap)}}_{i} - \begin{cases}
        			\sum\limits_{j \in \mathcal{Q}_{n}} a_{i j}(\tilde{\bm{x}}_{i} - \tilde{\bm{x}}_{j}), & {\text{if } i \in \mathcal{Q}};  \\
        			0, & {\text{if } i \in \mathcal{P}},
        		\end{cases}  
        		\label{Eq: CA-general-rb}
        	\end{align}
        \end{subequations}
        where $\tilde{\bm{u}}^{\rm{(adap)}}_{i} \in \mathbb{R}^{3}$ with each element is expressed as 
        \begin{equation} \label{Eq: CA-u3-re}
        	\tilde{u}^{\rm{(adap)}}_{i, \ell} =  c_{i} \cdot
        	\begin{cases}
        		\varepsilon_{i, \ell} \, {\rm sgn} \left(\tilde{x}_{i, \ell}\right), & {\text{if  }} |\tilde{x}_{i, \ell}| > \varepsilon_{i, \ell}; \\
        		\tilde{x}_{i, \ell}, & {\text{if  }} |\tilde{x}_{i, \ell}| \leq \varepsilon_{i, \ell},
        	\end{cases} 
        \end{equation}
        with $\ell \in \{1, 2, 3\}$, and the variables above have similar meanings to those in~\eqref{Eq: CA-general}, \eqref{Eq: CA-Pu}, and \eqref{Eq: CA-Qu}. Before we present the main result of this section, two preliminary assumptions in stability theory are reproduced below.
        
        \begin{assumption}[Strongly connected] \label{Asump: graph}
        	The graph $\mathcal{G}$ regarding the system given by $\mathcal{Q}_{n}$, $n = 1, \cdots, N$, is strongly connected, i.e., there is a path in each direction between each pair of vertices of the graph.
        \end{assumption}
        It is evident that a control strategy cannot be achieved if an isolated node does not receive information from any other nodes. As a result, this strong-connection assumption holds for the ABS swarm under study.
        
        \begin{assumption}[Initial position]  \label{Asump: initial}
        	At $t = 0$, the ABSs are outside mutual collision regions.
        \end{assumption}
        It is apparent that Assumption~\ref{Asump: initial} holds in practice if the initial positions of ABSs are well chosen.
        
        We are now in a position to present the main result of this section. The feasibility of the ABS swarm control strategy with collision avoidance can be determined by the following theorem.        
        \begin{theorem} \label{Thm: control-CA}
        	Given Assumptions \ref{Asump: graph} and \ref{Asump: initial}, the ABS swarms with dynamics given by~\eqref{Eq: CA-system} can reach the target positions while maintaining collision avoidance.
        \end{theorem}
        
        \begin{IEEEproof}
        	See Appendix \ref{Appendix: CA}.
        \end{IEEEproof}
        
        As a particular case, if there is only one ABS swarm performing mission execution, i.e., $M = 0$ and $N = 1$, the movement control strategy discussed above reduces to the case reported in our previous work \cite{Fan2023}.

\section{Simulation Results and Discussions} \label{sec:simulation} 
    In this section, the Monte-Carlo simulation results are presented and discussed. In the relevant simulation experiments, a wireless network with a square coverage area of length and width \SI{1}{km} is assumed. The patrol UAV flies at the height of \SI{25}{m} for detecting potential coverage holes. According to Technical Report 3GPP TR 23.754 \cite{3GPPTR23.754}, the flying height of UAV is limited to below \SI{300}{m}. Thus, if Config.~1 is applied, a single ABS flies at the height of \SI{150}{m}; and if Config.~2 is used, one of the ABS swarm flies at the height of \SI{300}{m} while the other three ABSs construct a regular tetrahedron at the height of \SI{150}{m}. 
    
    Also, we assume the received SINR threshold $\gamma_{\rm{th}} = \SI{11.3}{dB}$. That is, a received SINR higher than \SI{11.3}{dB} is assumed to be sufficient for effective communications, whereas the SINR level lower than \SI{11.3}{dB} is considered weak and causes a coverage hole. Moreover, the exclusion radius of checkpoints ($D$) is the same as the average distance between terrestrial BSs. For brevity, the main parameters used in the pertaining simulation experiments are summarized in Table~\ref{Tab: ParameterSetting}. 
    
    \begin{table}[t!]
    	\centering
    	\renewcommand\arraystretch{1.25}
    	\caption{Parameter Setting of Simulation Experiments.}
    	\renewcommand\arraystretch{1.2}
    	\footnotesize	
    	\begin{threeparttable}
    		\begin{tabular}{!{\vrule width0.12em} c | c !{\vrule width0.12em}}
    			\Xhline{0.12em} 
    			\hspace{20pt}{\bf Parameter (Symbol)} \hspace{20pt} &  \hspace{20pt} {\bf Value} \hspace{20pt} \\
    			\Xhline{0.12em} 
    			Density of terrestrial BSs ($\lambda_{B}$) & \SI{100}{BSs/km^2} \\
    			\hline
    			Bandwidth of terrestrial BSs  & \SI{15}{MHz} \\
    			\hline
    			Height of terrestrial BSs &  \SI{30}{m} \cite{Al-Ahmed2020Optimal} \\
    			\hline
    			Tx power of terrestrial BSs ($P_{{\rm BS}}^{\rm{tx}}$) & \SI{43}{dBm} \cite{Jahid2020Techno-Economic} \\
    			\hline
    			Nakagami shape factor ($m$) & 3 \cite{Guo2022Coverage}\\
    			\hline
    			Path-loss exponent ($\alpha_{\rm{B}}, \alpha_{\rm{A}, L}, \alpha_{\rm{A}, N}$) & $2.2, 2, 2.5$ \cite{Guo2022Coverage} \\
    			\hline
    			Non-/Los environment variable ($a, b$) & $\pi/18, 0.11$ \cite{Qin2024Coverage} \\
    			\hline
    			Height of the patrol UAV &  \SI{25}{m} \\
    			\hline
    			Received SINR threshold ($\gamma_{\rm{th}}$) & \SI{11.3}{dB} \\
    			\hline
    			Density of checkpoints ($\lambda_{\rm{CP}}$) & \SI{50}{points/km^2} (default) \\
    			\hline
    			Exclusion radius of checkpoints ($D$) & \SI{100}{m} \\
    			\hline
    			Tx power of the ABSs ($P^{\rm{tx}}_{\rm{ABS}}$) & \SI{37}{dBm} \cite{Ozbaltan2025Three-Dimensional}\\
    			\Xhline{0.12em} 
    		\end{tabular}
    		\label{Tab: ParameterSetting}
    	\end{threeparttable}
    \end{table}

    \subsection{Effectiveness of Coverage Hole Detection and Recovery} \label{Section-VA}
        
\begin{figure}[t]
        	\centering
        	\includegraphics[width=0.85\linewidth]{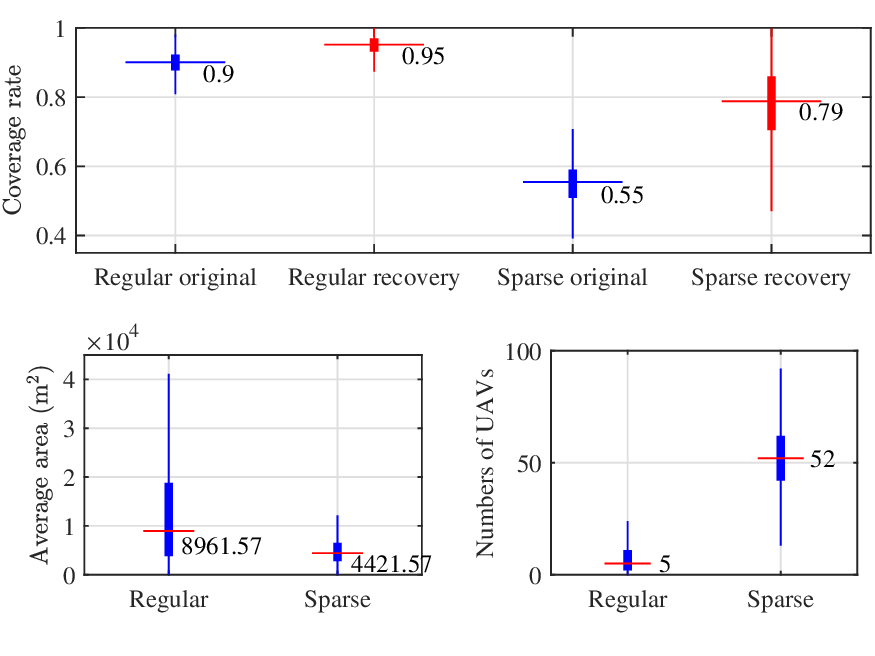}
        	\caption{The coverage rate (upper), the average improved coverage area per ABS (left lower), and the number of ABSs used in recovery (right lower) in both {\it regular} and {\it sparse} cases.} 
        	\label{Fig-7}
        \end{figure}

        To evaluate the effectiveness of our coverage hole detection and recovery method, we perform Monte Carlo simulations with 2000 independent trials. Box plots are employed to capture the statistical variability in the results. Fig.~\ref{Fig-7} presents the simulation outcomes for two scenarios: {\it regular} and {\it sparse}. The {\it regular} scenario corresponds to terrestrial BS deployments achieving more than $80\%$ coverage in the area of interest, whereas the {\it sparse} scenario refers to cases where only $30\%$ of BSs remain operational. The default checkpoint density is set to $\lambda_{\rm CP} = 50$ \si{points/km^2}.

        The upper panel of Fig.~\ref{Fig-7} compares the original coverage rate with the coverage rate after recovery in both scenarios. In the regular case, the median coverage rate increases from $0.90$ to $0.95$, whereas in the sparse case it increases from $0.55$ to $0.79$. These results indicate that our method enhances area coverage in both scenarios, with more pronounced improvement under sparse conditions.

        The lower panel of Fig.~\ref{Fig-7} shows the average coverage improvement per ABS and the total number of ABSs deployed for recovery. The median improved coverage per ABS in the regular scenario is $8961.57$ \si{m^2}, which is more than twice that in the sparse scenario ($4421.57$ \si{m^2}). Conversely, the median number of ABSs used for recovery in the regular case is $5$, which is less than one-tenth of the $52$ ABSs required in the sparse scenario. This discrepancy arises because the sparse scenario exhibits a lower average SINR, necessitating more ABSs to achieve adequate coverage. In practice, deploying multiple ABS swarms is more efficient than individual ABSs for sparse scenarios, as a single ABS provides relatively limited coverage. Finally, theoretical calculations based on \eqref{Eq:num-aBs-new} suggest a feasible number of ABSs in the range $[3, 80]$, which aligns closely with the simulation results shown in the lower panel of Fig.~\ref{Fig-7}.
       
        \begin{figure}[t]
        	\centering
        	\includegraphics[width=0.85\linewidth]{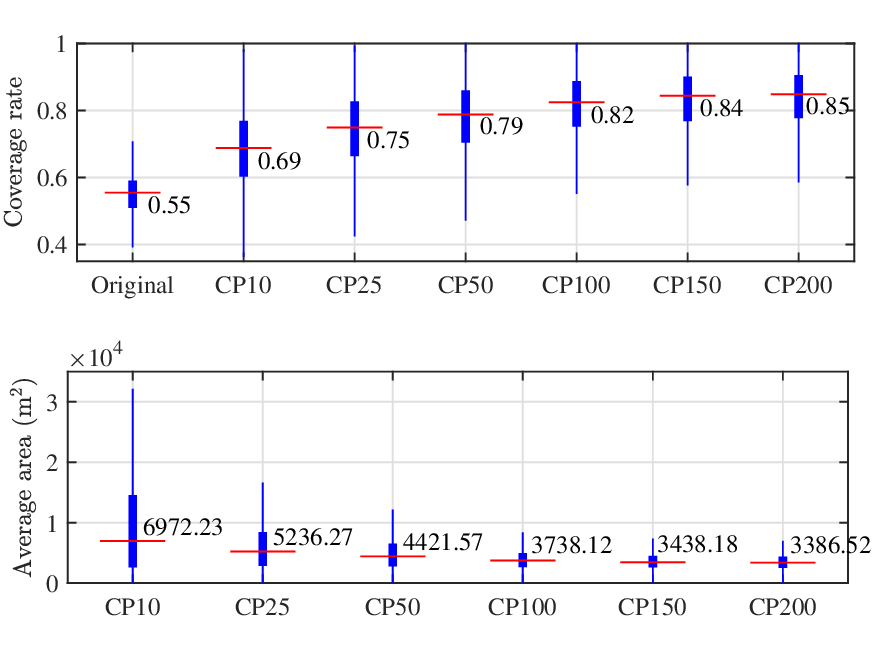}
        	\caption{The coverage rate (upper) and the average improved coverage area per ABS (lower) with $\lambda_{\rm{CP}} = \{10, 25, 50, 100, 150, 200\}$ \si{points/km^2} in {\it sparse} case.} 
        	\label{Fig-8}
        \end{figure}
        
        Fig.~\ref{Fig-8} presents the simulation results for the sparse scenario, showing both the coverage rate and the average coverage improvement per ABS for various checkpoint densities $\lambda_{\rm CP} = \{10, 25, 50, 100, 150, 200\}$ \si{points/km^2}, labeled as ``CP10", ``CP25", ``CP50", ``CP100", ``CP150", and ``CP200", respectively. 

        The upper panel of Fig.~\ref{Fig-8} compares the original coverage rate with that after recovery. As expected, the coverage rate increases with checkpoint density, but the incremental gain diminishes as the number of checkpoints becomes large. In particular, when $\lambda_{\rm CP} = 50$ \si{points/km^2}—approximately half of the terrestrial BS density $\lambda_B$—the coverage rate after recovery reaches $79\%$, which is nearly identical to the {\it regular} scenario ($80\%$). This observation indicates that when the number of checkpoints approaches the minimum required to replace out-of-service BSs, the recovered coverage can match the performance of the regular network.

        The lower panel of Fig.~\ref{Fig-8} shows that the average coverage improvement per ABS decreases with increasing checkpoint density. This trend occurs because, although more ABSs are deployed as checkpoint density increases, the marginal improvement in coverage diminishes once the density is sufficiently high, as anticipated.
    
    \subsection{Comparison of Proposed, BS Location-based, and Grid-based Methods}

\begin{figure}[t]
			\centering
			\includegraphics[width=1.0\linewidth]{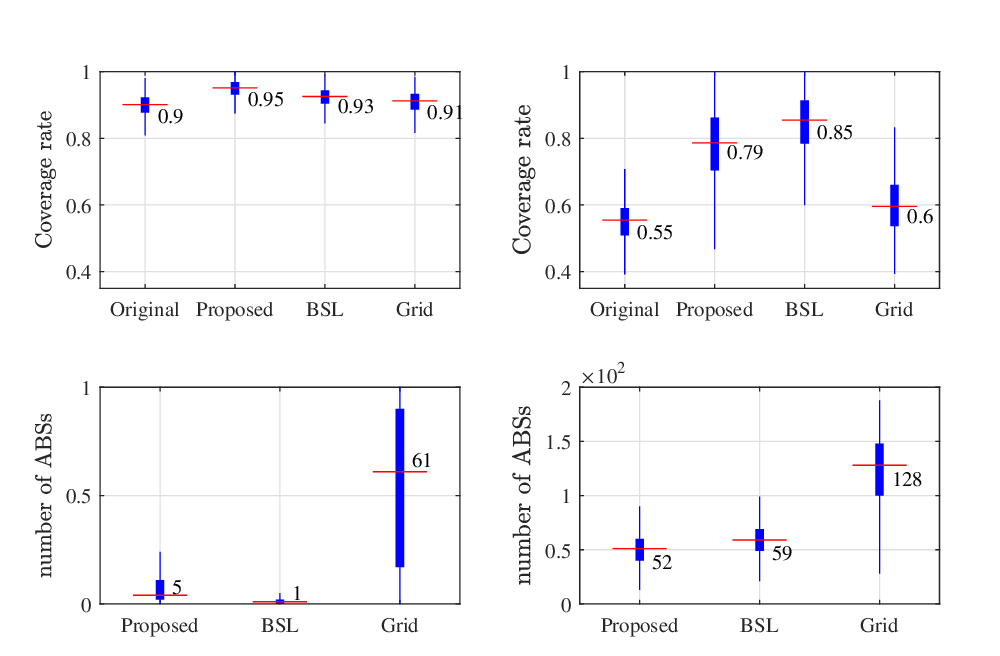}
			\caption{Comparison of the coverage rate (top) and the number of ABSs for recovery (bottom) among the proposed, BS location-based, and grid-based methods in the \textit{regular} (left) and \textit{sparse} (right) cases.} 
			\label{Fig-9}
		\end{figure}

		Fig.~\ref{Fig-9} presents a comparative evaluation of the coverage rate (top) and the number of ABSs required for recovery among the proposed method, BS location-based (BSL) method, and grid-based (Grid) method, with the original network state (Original) serving as a baseline. Specifically:  
		
		\begin{itemize}
		    \item \textbf{Original (Baseline)}: the unmodified network state obtained from the raw data, with no ABS deployment; used as a reference.  
		    \item \textbf{Proposed}: the method introduced in this work, formalized in Algorithm~\ref{Alg:CHD-RT}.  
		    \item \textbf{BSL}: a variant that uses the locations of terrestrial BSs as prior knowledge and configures Algorithm~\ref{Alg:CHD-RT} with these locations as checkpoints.  
		    \item \textbf{Grid}: a baseline that places checkpoints at the centers of a uniform $7 \times 7$ grid and employs a serpentine traversal from top to bottom to ensure complete coverage.
		\end{itemize}
		
		In the standard scenario (left panels of Fig.~\ref{Fig-9}), all three methods achieve similarly high coverage rates ($0.93$-$0.95$), which are close to the original coverage rate ($0.90$). However, the proposed and BSL methods require far fewer ABSs ($5$ and $1$, respectively) compared to the grid-based method ($61$), demonstrating superior deployment efficiency.  
		
		In the sparse scenario (right panels of Fig.~\ref{Fig-9}), the proposed and BSL methods achieve higher coverage rates ($0.79$ and $0.85$) compared to the grid-based method ($0.60$), while deploying fewer ABSs ($52$ and $59$ versus $128$). These results indicate that both the proposed and BSL methods achieve comparable or higher coverage with substantially fewer ABSs, particularly under sparse network conditions. Unlike the grid-based approach, which uniformly distributes ABSs without accounting for the spatial pattern of coverage loss, the proposed method adaptively targets actual coverage deficits, thereby minimizing redundant deployments and improving resource efficiency.
		
		Although the BSL method can be viewed as a special case of the proposed approach, our method consistently delivers higher recovery rates in regular scenarios by reinforcing coverage around BSs. It achieves notable improvements in sparse scenarios via more comprehensive detection. Overall, these findings underscore the benefits of adaptive, demand-driven ABS deployment strategies for efficient and effective network restoration across diverse environments.

    \subsection{Case Study 1: Coverage Hole Detection and Recovery}

    To illustrate the effectiveness of the proposed method, we perform a case study of coverage hole detection and recovery under both {\it regular} and {\it sparse} terrestrial BS deployments.  
	
	\begin{figure}[t]
		\centering
		\captionsetup[subfigure]{margin=2pt}
		\subfloat[Patrol UAV trajectory and detected coverage holes.]{
			\label{Fig-10a}
			\includegraphics[width=0.85\linewidth]{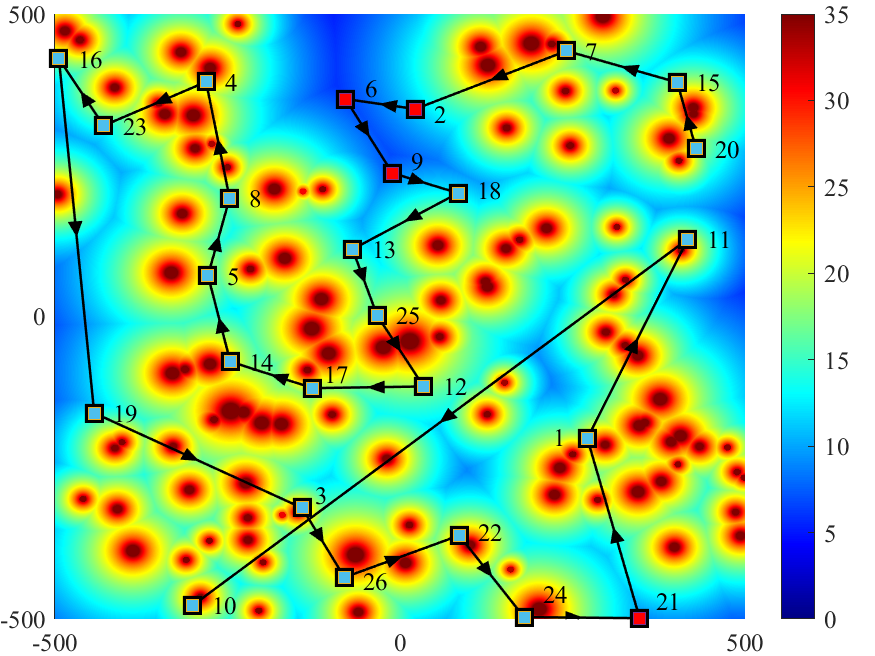}
		} \\
		\subfloat[SINR heat map after recovery of coverage holes.]{
			\label{Fig-10b}
			\includegraphics[width=0.85\linewidth]{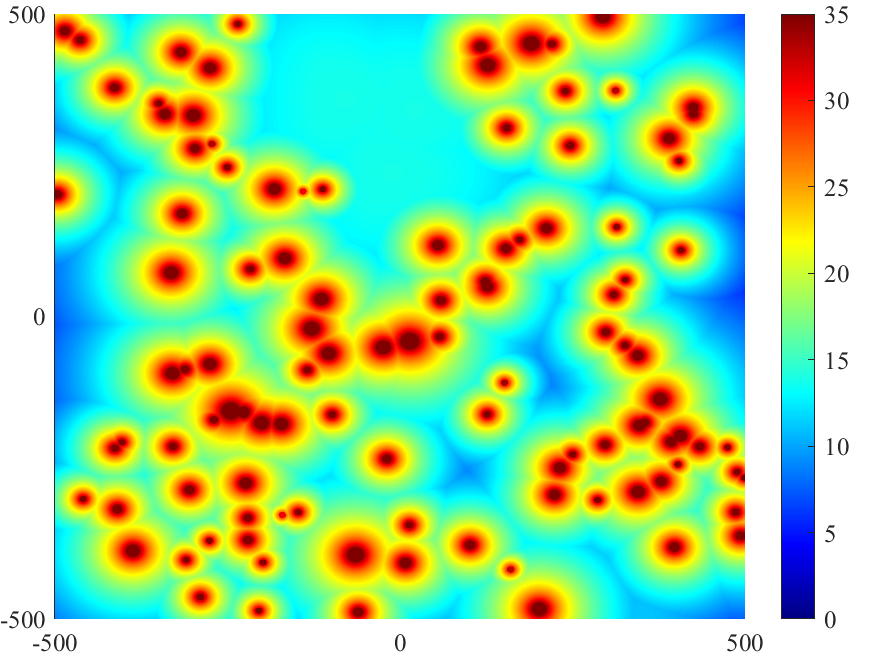}
		}
		\caption{Simulation results of coverage hole detection and recovery for {\it regular} terrestrial BSs.} 
		\label{Fig-10}
	\end{figure}
	
	\subsubsection{\underline{Regular Terrestrial BSs}}
	
	Fig.~\ref{Fig-10} shows the simulation results for the {\it regular} case. The background of Fig.~\ref{Fig-10a} depicts the SINR heat map, where $12.52\%$ of the area experiences SINR below \SI{11.3}{dB}. We generate $26$ checkpoints via a Mat\'{e}rn hard-core process with exclusion radius $D = \SI{100}{m}$. The shortest path for the patrol UAV to traverse these checkpoints is obtained using Dijkstra's algorithm, as shown by the arcs in Fig.~\ref{Fig-10a}.  
	
	The detection results indicate that $4$ out of $26$ checkpoints fall within coverage holes, consistent with the SINR statistics (approximately $15.38\%$ below the threshold). Fig.~\ref{Fig-10b} depicts the SINR heat map after recovery, showing significant improvements around checkpoints 2, 6, 9, and 21. At a per-square-meter granularity, the overall coverage rate increases by approximately $5.8\%$, demonstrating the method’s effectiveness in regular networks.
	
	\subsubsection{\underline{Sparse Terrestrial BSs}}
	
\begin{figure}[t]
		\centering
		\captionsetup[subfigure]{margin=2pt}
		\subfloat[Patrol UAV trajectory and detected coverage holes.]{
			\label{Fig-11a}
			\includegraphics[width=0.85\linewidth]{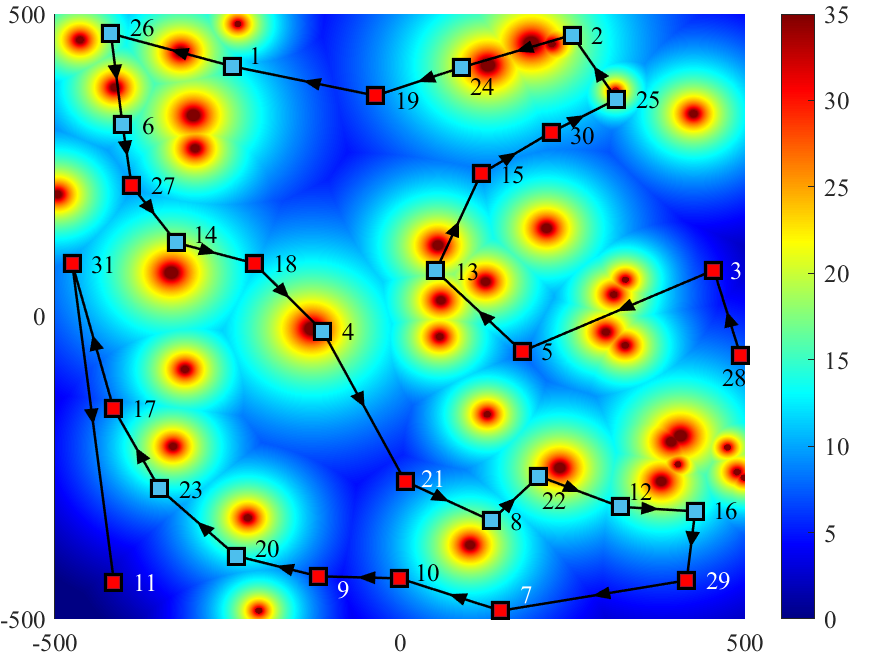}
		} \\
		\subfloat[SINR heat map after recovery of coverage holes.]{
			\label{Fig-11b}
			\includegraphics[width=0.85\linewidth]{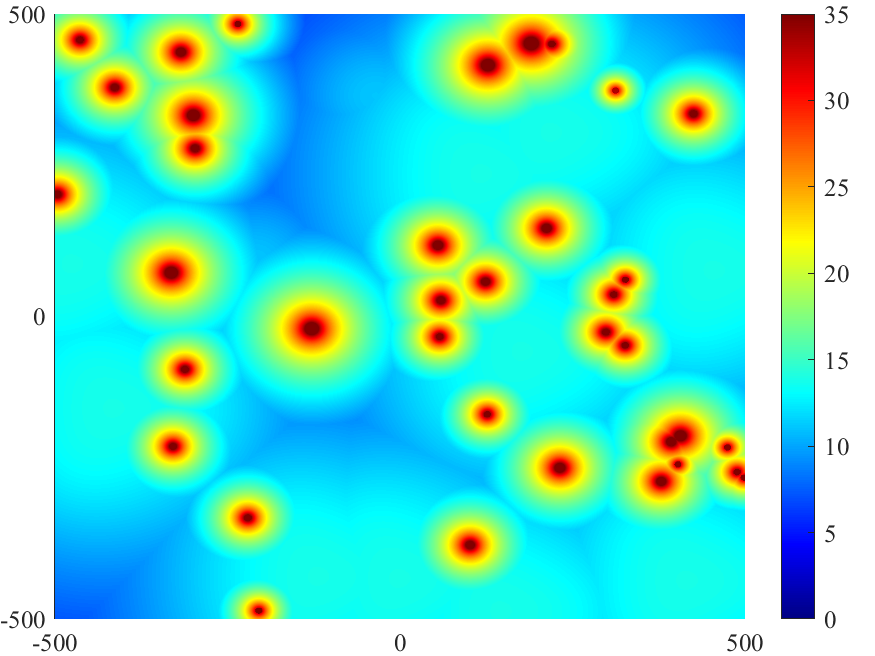}
		}
		\caption{Simulation results of coverage hole detection and recovery for {\it sparse} terrestrial BSs.}
		\label{Fig-11}
	\end{figure}

	In the case of network emergencies, such as earthquakes or hurricanes, most terrestrial BSs may fail. Fig.~\ref{Fig-11} illustrates coverage detection and recovery when only $30\%$ of BSs remain functional. Fig.~\ref{Fig-11a} shows that only $37$ out of $110$ BSs are active, leaving $43.42\%$ of the area below \SI{11.3}{dB}. The patrol UAV’s $31$ preset checkpoints, and the shortest path derived by Dijkstra’s algorithm are also shown. Of these, only $15$ checkpoints lie within active coverage areas.  
	
	After deploying ABS swarms, the SINR heat map in Fig.~\ref{Fig-11b} becomes more uniform, increasing the coverage rate from $56.58\%$ to $81.90\%$ using $52$ ABSs to replace the $73$ failed BSs. Comparing Fig.~\ref{Fig-10b} and Fig.~\ref{Fig-11b} reveals residual areas with suboptimal coverage, even in the regular case. In practice, increasing checkpoint density or the number/frequency of patrol UAVs can further enhance coverage.
	
	\subsubsection{\underline{Effect of Patrol Visiting Order}}
	
	To examine the impact of different checkpoint visiting orders, we simulate six variations using the same checkpoint graph from Fig.~\ref{Fig-11a} in a sparse scenario with an original coverage rate of $56.58\%$. Detection is initiated from four far-end points (left-most: 31, right-most: 28, top-most: 26, bottom-most: 7) and two random inner points (15 and 4).  
	
\begin{figure*}[!ht]
    	\begin{minipage}{.74\linewidth}
    		\centering
    		\renewcommand\arraystretch{1.25}
    		\tabcaption{Statistics of different visiting orders on the same checkpoint graph.}
    		\label{Tab: Statistics} 
    		\renewcommand\arraystretch{1.2}
    		\footnotesize	
    		\begin{threeparttable}
    			\begin{tabular}{ !{\vrule width 0.12em} c !{\vrule width 0.12em} c|c|c|c !{\vrule width 0.12em}}
    				\Xhline{0.12em} 
    				\multirow{2}{*}{\bf Starting Point}  & \multirow{2}{*}{\bf  \# of ABSs} &  \multirow{2}*{\shortstack{{\bf Coverage Rate} \\ {\bf After Recovery}}}  &  \multicolumn{2}{c !{\vrule width1.2pt}}{\bf ABS Configurations} \\ 
    				\Xcline{4-5} {0.08em} 
    				& & & {\bf Config. 1} & {\bf Config. 2}   \\
    				\Xhline{0.12em} 
    				\makecell{The left-most \\ point 31}  & $46$  & $80.25\%$  
    				& \makecell{\{5\}, \{7\}, \{15\}, \\ \{18\}, \{19\}, \{30\}} 
    				& \makecell{\{3, 28, 29\}, \{9, 10, 21\}, \\ \{11, 17\}, \{27, 31\}} \\
    				\hline
    				\makecell{The right-most \\ point 28} & $52$  & $81.90\%$ 
    				& \makecell{\{18\}, \{19\}, \\ \{21\}, \{27\}} 
    				& \makecell{\{3, 5, 28\}, \{7, 9, 10, 29\}, \\ \{15, 30\},  \{11, 17, 31\}}  \\
    				\hline
    				\makecell{The top-most \\ point 26} & $55$  & $85.06\%$ 
    				&  \{18\}, \{19\}, \{27\} 
    				& \makecell{\{3, 5, 11, 21, 28, 29\}, \\ \{7, 9, 10\}, \{15, 30\}, \{17, 31\}}  \\
    				\hline
    				\makecell{The bottom-most \\ point 7} & $52$  & $84.02\%$  
    				& \makecell{ \{7\}, \{18\}, \\ \{19\}, \{27\}} 
    				& \makecell{\{3, 5, 28, 29\}, \{9, 10\}, \\ \{11, 17, 31\}, \{15, 30\}}  \\
    				\hline
    				Random point 15 & $52$  & $83.76\%$ 
    				& \makecell{\{5\}, \{18\}, \\ \{19\}, \{27\}} 
    				& \makecell{\{3, 28, 29\}, \{7, 11, 17, 31\} \\ \{9, 10, 21\}, \{15, 30\}}   \\
    				\hline
    				Random point 4 & $46$  & $80.66\%$  
    				& \makecell{\{5\}, \{15\}, \{18\}, \\ \{19\}, \{27\}, \{30\}} 
    				& \makecell{\{3, 28, 29\}, \\ \{7, 11, 17, 31\}, \{9, 10, 21\}}   \\
    				\Xhline{0.12em} 
    			\end{tabular}
    		\end{threeparttable}
    	\end{minipage}
    	\begin{minipage}{0.25\linewidth}
    		\centering
    		\vspace{30pt}
    		\includegraphics[width=0.8\linewidth]{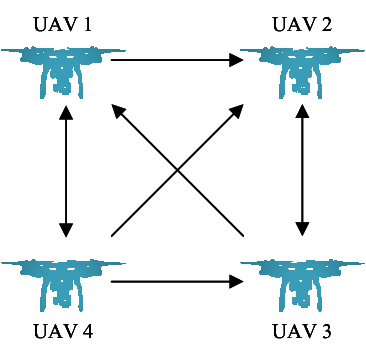}
    		\figcaption{A communication model among ABSs in the same swarm.}
    		\label{Fig-12}
    	\end{minipage}
    \end{figure*}

	Table~\ref{Tab: Statistics} summarizes the results. The total number of ABSs used and the post-recovery coverage rate remain similar across different starting points. However, the visiting order significantly affects ABS configurations. For instance, starting from the left-most point 31 requires six individual ABSs to recover areas around checkpoints 5, 7, 15, 18, 19, and 30, and ten ABS swarms to jointly serve checkpoints $\{3, 28, 29\}$, $\{9, 10, 21\}$, $\{11, 17\}$, and $\{27, 31\}$. By contrast, starting from the right-most point 28 requires four individual ABSs and twelve ABS swarms for the corresponding checkpoint clusters.
  
\subsection{Case Study 2: Multi-ABS Control with Collision Avoidance} 

To demonstrate the effectiveness of the proposed control strategies \eqref{Eq: CA-Pu} and \eqref{Eq: CA-Qu} for geometric formation and collision avoidance in ABS swarms, we consider the first nine checkpoints in the right-upper quarter of Fig.~\ref{Fig-11} (labeled $28$, $3$, $5$, $13$, $15$, $30$, $25$, $2$, and $24$). In this scenario, five ABS swarms are scheduled to recover coverage holes around checkpoints $28$, $3$, $5$, $15$, and $30$, respectively.

The simulation parameters are set as follows: collision radius $r_{c} = \SI{10}{m}$, communication radius $r_{d} = \SI{30}{m}$, maximum ABS speed $20$ \si{m/s}, and ABS mass $1$ \si{kg}. The control coefficients are $k_1 = 0.01$ \si{kg/s}, $k_2 = 0.02$ \si{kg/m}, with $b_i = 0.5$ and $\varepsilon_{i, \ell} = 10$ for all $\ell \in \{1,2,3\}$.  

Each ABS has a 3D position vector $\bm{x}_i = [x_{iX}, x_{iY}, x_{iZ}]^{\rm T}$ in the inertial coordinate system, following the right-hand rule. The interconnection topology of each swarm is depicted in Fig.~\ref{Fig-12}, and the corresponding weighted adjacency matrix $\bm{A}$ used in \eqref{Eq: CA-Qu} is
\begin{equation}
	\bm{A} = 0.01 \times 
	\begin{bmatrix}
		0 & 1 & 0 & 1 \\
		0 & 0 & 1 & 0 \\
		1 & 1 & 0 & 0 \\
		1 & 1 & 1 & 0
	\end{bmatrix}.
\end{equation}
The factor $0.01$ is a communication control gain to maintain input magnitudes in the same order. Unlike the leader-follower structure in \cite[Fig.~3]{Fan2023}, this leaderless interconnection improves robustness against failures.

\begin{figure*}[t]
	\centering
	\includegraphics[width=0.97\linewidth]{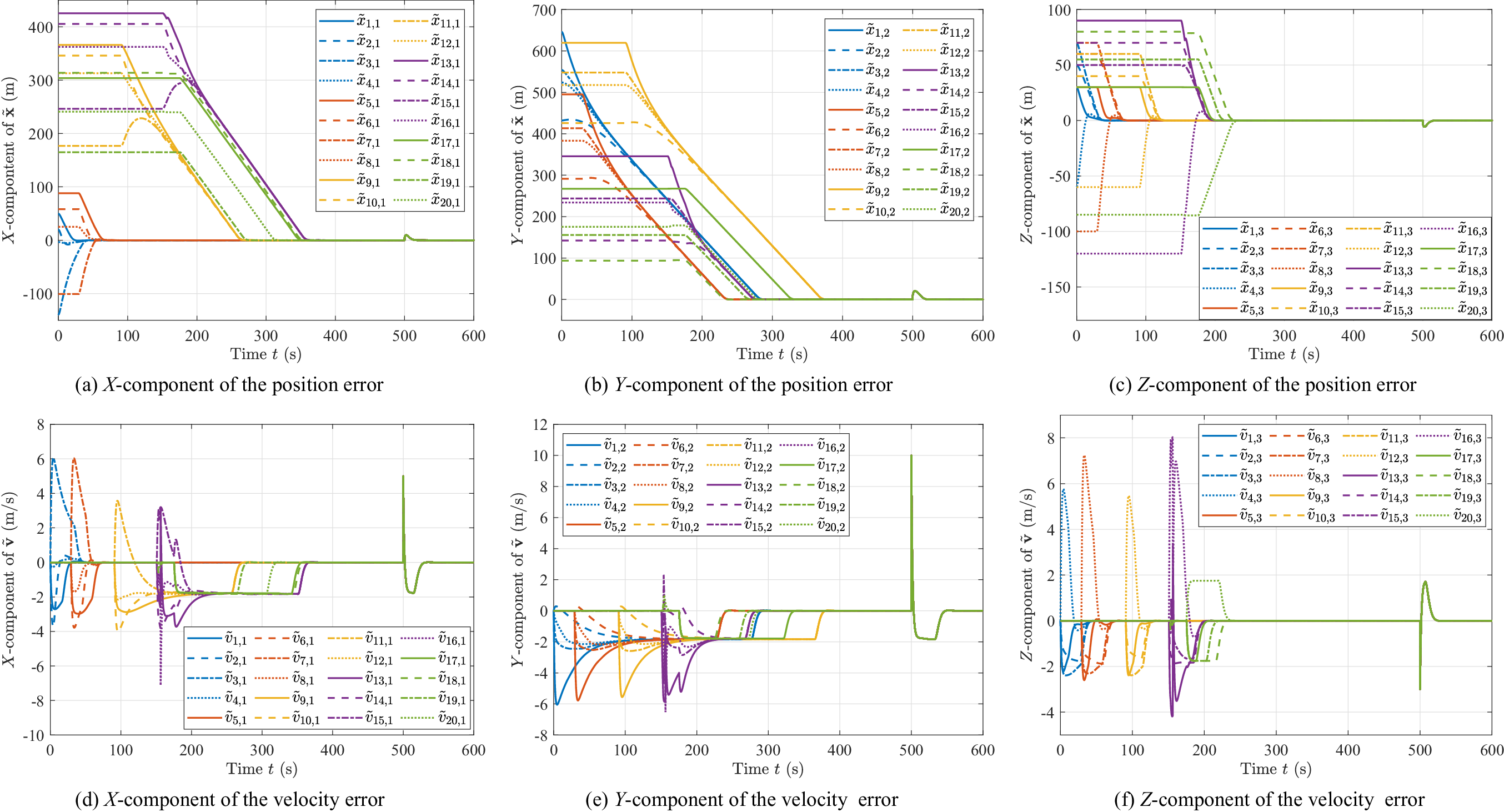}
	\caption{Error dynamics of ABS swarms for recovery of five checkpoints. The upper panel shows position errors, and the lower panel shows velocity errors.}
	\label{Fig-13}
\end{figure*}

\begin{figure}[!t]
	\centering
	\captionsetup[subfigure]{margin=2pt}
	\subfloat[2D trajectories of ABS swarms.]{
		\label{Fig-14a}
		\includegraphics[width=0.75\linewidth]{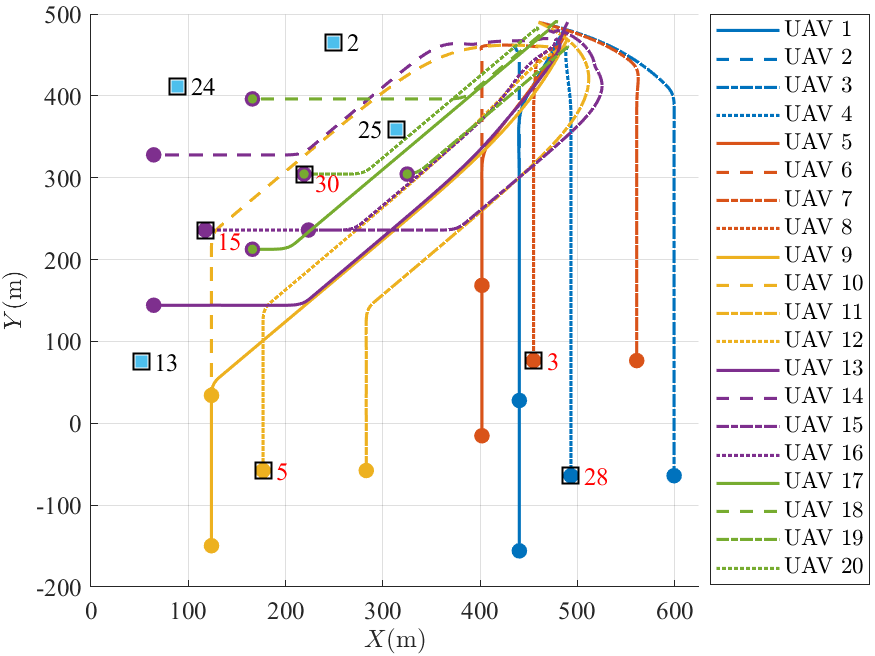}} \\
	\subfloat[Minimum distance between any two ABSs.]{
		\label{Fig-14b}
		\includegraphics[width=0.75\linewidth]{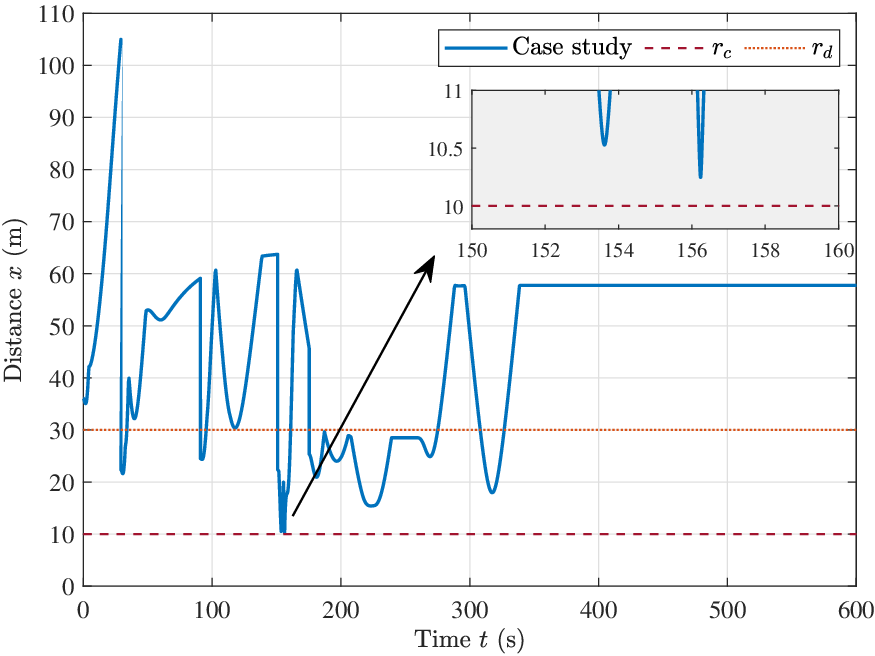}}
	\caption{2D trajectories and collision avoidance of ABS swarms. Each ABS takes off from a randomly selected location, forms a stable swarm, and flies to the target checkpoints. The minimum inter-ABS distance remains above the preset collision radius $r_c = \SI{10}{m}$ throughout the mission.}
	\label{Fig: case-ca}
\end{figure}

Fig.~\ref{Fig-13} illustrates the error dynamics of the ABS swarms during the recovery. Each component of the position and velocity errors converges to zero within $400$ \si{s}, with faster convergence for higher ABS speeds. Around $t = 500$ \si{s}, slight jitters appear due to an impulse environmental disturbance $[5, 10, -3]^{\rm T}$ simulating gusts, after which the errors quickly return to zero. These results demonstrate that the swarms maintain formation, effectively reach target locations, and are robust to disturbances.

For intuitive visualization, Fig.~\ref{Fig-14a} displays the 2D trajectories of the five ABS swarms as they recover the target checkpoints. Each swarm forms after takeoff and proceeds to the predetermined checkpoint locations. Fig.~\ref{Fig-14b} depicts the minimum distance between any two ABSs, confirming that it always remains above $r_c = \SI{10}{m}$, ensuring collision-free operation.

\section{Concluding Remarks} \label{sec:conclusion}
    This paper introduced a novel UAV-based wireless network framework for detecting and recovering terrestrial coverage holes, leveraging space–air–ground resources to enable fast and resilient network restoration. An online scheduling scheme was developed to identify coverage gaps and determine the number of ABSs required based on gap size, while a collision-aware movement control strategy was designed for both single-ABS and ABS-swarm deployments. Simulation results demonstrated that the proposed approach effectively restores coverage, even under sparse terrestrial BS conditions, benefiting from CoMP transmissions in ABS swarms. Beyond recovery, the framework can enhance overall coverage. 
    
	From an operational perspective, several practical factors are not explicitly modeled in the current work, but are critical for real-world deployments. In particular, UAV energy consumption and endurance constraints will limit the feasible patrol duration and recovery frequency, and the availability and capacity of backhaul connectivity (e.g., via satellite or terrestrial gateways) will affect the rate at which control information and user traffic can be relayed.  In future work, these aspects will be incorporated into the framework by developing energy-aware scheduling policies, endurance-constrained patrol planning, and backhaul-aware resource allocation, which are especially relevant for large-scale non-terrestrial network deployments.

\appendices

\section{Proof of Theorem~\ref{Thm:num-aBS}} \label{Appendix:num} 

Consider a given area $\mathcal{V} \subset \mathbb{R}^{2}$ and an effective coverage radius $R$ for the ABSs. Suppose that $\mathcal{V}$ can be seamlessly covered by $N_{c} (R)$ ABSs, i.e.,
\begin{equation}
	\mathcal{V} \subset \bigcup_{i = 1}^{N_{c}(R)} \mathcal{B}({\bm x}_{i}, R),
\end{equation}
where $\mathcal{B}({\bm x}_{i}, R)$ denotes a ball of radius $R$ centered at ${\bm x}_{i}$. Then, it follows that
\begin{align*}
	{\rm{vol}} (\mathcal{V}) \leq {\rm{vol}} \left(\bigcup_{i = 1}^{N_{c} (R)} \mathcal{B} \left({\bm x}_{i}, R  \right)\right) \leq \sum_{i=1}^{N_{c} (R)} {\rm{vol}} \left(\mathcal{B} \left({\bm x}_{i}, R\right)\right) \\ = N_{c} (R) \, {\rm{vol}} \left(\mathcal{B} \left(\bm{o}, R\right)\right).
\end{align*}
Dividing both sides by ${\rm vol}(\mathcal{B}(\bm{o}, R))$ yields the lower bound:
\begin{equation}
	\frac{{\rm vol}(\mathcal{V})}{{\rm vol}(\mathcal{B}(\bm{o}, R))} \leq N_{c}(R).
\end{equation}

For the upper bound, consider $N_{p}(R)$ disjoint balls of radius $R/2$ centered at ${\bm y}_{i} \in \mathcal{V}$. Although these balls may not lie entirely inside $\mathcal{V}$, they fit within the inflated set
\begin{equation}
	\bigcup_{i=1}^{N_{p}(R)} \mathcal{B}\left({\bm y}_{i}, R/2\right) \subset \mathcal{V} \oplus \mathcal{B}(\bm{o}, R/2),
\end{equation}
where $\oplus$ denotes the Minkowski sum. Since the balls are disjoint, the distance between any two centers is at least $R$. Taking the volume on both sides gives
	\begin{align*}
		{\rm{vol}} \left(\mathcal{V} \oplus \mathcal{B} \left(\bm{o}, R/2\right)\right) 
		& \geq {\rm{vol}} \left(\bigcup_{i = 1}^{N_{p} (R)} \mathcal{B} \left({\bm y}_{i}, R/2\right)\right) \\
		& = N_{p} (R) \, {\rm{vol}} \left(\mathcal{B} \left(\bm{o}, R/2\right)\right),
	\end{align*}
which implies
\begin{equation}
	N_{p}(R) \leq \frac{{\rm vol}\Big(\mathcal{V} \oplus \mathcal{B}(\bm{o}, R/2)\Big)}{{\rm vol}\Big(\mathcal{B}(\bm{o}, R/2)\Big)}.
\end{equation}

Finally, by \cite[Lemma 4.2.8]{Vershynin2018}, we have $N_{c}(R) \leq N_{p}(R)$. Combining the lower and upper bounds yields the desired result of Theorem~\ref{Thm:num-aBS}.

\section{The proof of Theorem~\ref{Thm: control-CA}} \label{Appendix: CA}
    Construct the following Lyapunov functional candidate
	\begin{equation} \label{eq: CA-V}
		V = V_{1} + V_{2} + V_{3} + V_{4},
	\end{equation}
    where 
	\begin{subequations}
		\begin{align}
			V_{1} & = \frac{1}{2} \sum_{\ell = 1}^{3} \sum_{n = 1}^{N} \sum_{i \in \mathcal{Q}_{n}} \sum_{j \in  \mathcal{Q}_{n}} w_{ij} \left(\tilde{x}_{i, \ell} - \tilde{x}_{j, \ell} \right)^{2},
			\label{eq: CA-V1} \\
			V_{2} & = \sum_{\ell = 1}^{3} \left(\sum_{i \in \mathcal{P}}\tilde{v}_{i, \ell}^{2} + \sum_{n = 1}^{N}  \sum_{i \in \mathcal{Q}_{n}} \tilde{v}_{i, \ell}^{2}\right),  
			\label{eq: CA-V2} \\
			V_{3} & = \sum_{\ell = 1}^{3} \sum_{k \in \mathcal{P} \cup \mathcal{Q}} \left(\sum_{i \in \mathcal{P}}  V^{(i, k)}_{\ell} \left({x}_{i, \ell}, {x}_{k, \ell}\right) \right. \nonumber \\
			& \quad \left. + \sum_{n = 1}^{N} \sum_{i \in \mathcal{Q}_{n}} V^{(i, k)}_{\ell} \left({x}_{i, \ell},  {x}_{k, \ell}\right)\right),
			\label{eq: CA-V3} \\
			V_{4} & = \sum_{\ell = 1}^{3} \left(\sum_{i \in \mathcal{P}} g_{i} (\tilde{x}_{i, \ell}) + \sum_{n =  1}^{N} \sum_{i \in \mathcal{Q}_{n}} g_{i} (\tilde{x}_{i, \ell})\right),
			\label{eq: CA-V4}
		\end{align} 
	\end{subequations}  
    with $w_{ij} = \left(a_{ij} + a_{ji}\right) / 2$ in \eqref{eq: CA-V1}, and the function $g_{i} (\tilde{x}_{i, \ell})$ with $\ell \in \{1, 2, 3\}$ in \eqref{eq: CA-V4} is expresses as 
	\begin{equation*}
		G_{i} (\tilde{x}_{i, \ell}) = c_{i} \cdot
		\begin{cases}
			\varepsilon_{i, \ell} |\tilde{x}_{i, \ell}| - \frac{\varepsilon_{i, \ell}^{2}}{2}, & {\rm{if  }}  |\tilde{x}_{i, \ell}| > \varepsilon_{i, \ell}; \\
			\frac{\tilde{x}_{i, \ell}^{2}}{2}, & {\rm{if  }} |\tilde{x}_{i, \ell}| \leq \varepsilon_{i, \ell}.
		\end{cases} 
    \end{equation*}

    First, differentiating $V_{1}$ with respect to time $t$ yields
	\begin{subequations}
		\begin{align} \label{eq: CA-V1-1}
			\dot{V}_{1} & = \sum_{\ell = 1}^{3} \sum_{n = 1}^{N} \sum_{i \in \mathcal{Q}_{n}} \sum_{j \in  \mathcal{Q}_{n}} w_{ij} \left(\tilde{x}_{i, \ell} - \tilde{x}_{j, \ell} \right) \left(\tilde{v}_{i, \ell} - \tilde{v}_{j, \ell}\right) \nonumber \\
			& \overset{(a)}{=}  2 \sum_{\ell = 1}^{3} \sum_{n = 1}^{N} \sum_{i \in \mathcal{Q}_{n}} \sum_{j \in  \mathcal{Q}_{n}} w_{ij} \left(\tilde{x}_{i, \ell} - \tilde{x}_{j, \ell} \right) \tilde{v}_{i, \ell}, 
		\end{align}
		{\rm {where the step $(a)$ follows $w_{ji} = w_{ij} \triangleq \left(a_{ij} + a_{ji}\right) / 2$.} {Next, take the time derivative of \eqref{eq: CA-V2}--\eqref{eq: CA-V4}, respectively, we have }}
		\begin{align} 
			\dot{V}_{2} & = \underbrace{2 \sum_{\ell = 1}^{3} \sum_{i \in \mathcal{P}} \tilde{v}_{i, \ell}  \dot{\tilde{v}}_{i, \ell}}_{\dot{V}_{2}^{(\mathcal{P})}} + \underbrace{2 \sum_{\ell = 1}^{3} \sum_{n = 1}^{N} \sum_{i \in \mathcal{Q}_{n}} \tilde{v}_{i, \ell} \dot{\tilde{v}}_{i, \ell}}_{\dot{V}_{2}^{(\mathcal{Q})}}, 
			\label{eq: CA-V2-1} \\
			\dot{V}_{3} & = \underbrace{2 \sum_{\ell = 1}^{3} \sum_{k \in \mathcal{P} \cup \mathcal{Q}} \sum_{i \in  \mathcal{P}} \frac{\partial V^{(i, k)}_{\ell} \left({x}_{i, \ell}, {x}_{k, \ell}\right)}{\partial {x}_{i, \ell}} \, \tilde{v}_{i, \ell}}_{\dot{V}_{3}^{(\mathcal{P})}} \nonumber\\
			& \quad + \underbrace{2\sum_{\ell = 1}^{3} \sum_{k \in \mathcal{P} \cup \mathcal{Q}} \sum_{n = 1}^{N}  \sum_{i \in \mathcal{Q}_{n}} \frac{\partial V^{(i, k)}_{\ell} \left({x}_{i, \ell}, {x}_{k, \ell}\right)}{\partial {x}_{i, \ell}} \, \tilde{v}_{i, \ell}}_{\dot{V}_{3}^{(\mathcal{Q})}},  
			\label{eq: CA-V3-1} \\
			\dot{V}_{4} & = \underbrace{\sum_{\ell = 1}^{3} \sum_{i \in \mathcal{P}} \dot{g}_{i} (\tilde{x}_{i,  \ell})}_{\dot{V}_{4}^{(\mathcal{P})}} + \underbrace{\sum_{\ell = 1}^{3} \sum_{n = 1}^{N} \sum_{i \in \mathcal{Q}_{n}} \dot{g}_{i} (\tilde{x}_{i, \ell})}_{\dot{V}_{4}^{(\mathcal{Q})}},
			\label{eq: CA-V4-1}  
		\end{align}
	\end{subequations}
    with $\dot{g}_{i} (\tilde{x}_{i, \ell}) =  c_{i} \, \varepsilon_{i, \ell} \, {\rm{sgn}}\left(\tilde{x}_{i, \ell}\right)\tilde{v}_{i, \ell}$ if $|\tilde{x}_{i, \ell}| > \varepsilon_{i, \ell}$; otherwise, $\dot{g}_{i} (\tilde{x}_{i, \ell}) =  c_{i} \, \tilde{x}_{i, \ell} \tilde{v}_{i, \ell}$. Then, combine \eqref{eq: CA-V} with \eqref{eq: CA-V1-1}--\eqref{eq: CA-V4-1}, the time derivative of $V$ satisfies 
	\begin{align}\label{eq: CA-V-1}
		\hspace{-1em} 
		\dot{V} & = \dot{V}_{1} + \dot{V}_{2} + \dot{V}_{3} + \dot{V}_{4} \nonumber\\
		& = \underbrace{\dot{V}_{1} + \dot{V}_{2}^{(\mathcal{Q})} + \dot{V}_{3}^{(\mathcal{Q})} + \dot{V}_{4}^{(\mathcal{Q})}}_{\dot{V}^{(\mathcal{Q})}} + \underbrace{\dot{V}_{2}^{(\mathcal{P})} + \dot{V}_{3}^{(\mathcal{P})} + \dot{V}_{4}^{(\mathcal{P})}}_{\dot{V}^{(\mathcal{P})}}. \hspace{-0.5em} 
	\end{align}
    Combine \eqref{Eq: CA-system} with \eqref{eq: CA-V1-1}--\eqref{eq: CA-V4-1}, we attain
	\begin{equation} \label{eq: CA-VQ}
		\dot{V}^{(\mathcal{Q})} =  - 2 \sum_{\ell = 1}^{3} \sum_{n = 1}^{N} \sum_{i \in \mathcal{Q}_{n}}  \left(b_{i} |\tilde{v}_{i, \ell}|^{2} - f_{R, \ell} \tilde{v}_{i, \ell}\right), 
	\end{equation}
    where the $f_{R, \ell}$ is the $\ell^{\rm{th}}$ element of $\bm{f}_{R} \in \mathbb{R}^{3}$ in \eqref{Eq: CA-general-fr}. Since $\tilde{v}_{i, \ell} = \dot{\tilde{x}}_{i, \ell} = \dot{x}_{i, \ell} - \dot{x}_{i, \ell}^{\rm{dest}} = v_{i, \ell}$, and recall \eqref{Eq: CA-general-fr}, we have
	\begin{align}\label{eq: CA-VP-1}
		\setcounter{equation}{22}
		\sum_{\ell = 1}^{3} \sum_{n = 1}^{N} \sum_{i \in \mathcal{Q}_{n}} f_{R, \ell} \tilde{v}_{i, \ell} 
		& = - \sum_{\ell = 1}^{3} \sum_{n = 1}^{N} \sum_{i \in \mathcal{Q}_{n}} \left(k_{1} |v_{i, \ell}|^{2} +  k_{2} |v_{i, \ell}|^{3}\right) \nonumber\\
		& \leq 0. 
	\end{align}
    By virtue of \eqref{eq: CA-VQ}--\eqref{eq: CA-VP-1}, it is obviously that $\dot{V}^{(\mathcal{P})} \leq 0$, and $\dot{V}^{(\mathcal{P})} = 0$ if and only if $\tilde{v}_{i, \ell} = 0$. Similarly, combine \eqref{Eq: CA-system} with \eqref{eq: CA-V2-1}--\eqref{eq: CA-V4-1}, we have  
	\begin{equation}\label{eq: CA-VP}
		\dot{V}^{(\mathcal{P})} = -2 \sum_{\ell = 1}^{3} \sum_{i \in \mathcal{Q}} \left(b_{i} |\tilde{v}_{i,  \ell}|^{2} - f_{R, \ell} \tilde{v}_{i, \ell}\right) \leq 0. 
	\end{equation}
    Combine \eqref{eq: CA-V-1} with \eqref{eq: CA-VQ}--\eqref{eq: CA-VP}, it is straightforward that $\dot{V} \leq 0$. Denote $\mathcal{M}$ as the largest invariance set $\mathcal{M} \triangleq \{\bm{x}_{1}, \cdots, \bm{x}_{M+4N}, \bm{v}_{1}, \cdots, \bm{v}_{M+4N} \mid \dot{V} \equiv 0\}$. Recall \eqref{eq: CA-V} and \eqref{eq: CA-V1-1}--\eqref{eq: CA-V4-1}, we have $\tilde{x}_{i, \ell} = 0$ and $\tilde{v}_{i, \ell} = 0$ when $\dot{V} \equiv 0$. Use LaSalle's invariance principle \cite[Thm. 3.5]{Slotine1990}, we have $\lim_{t \rightarrow + \infty} |\tilde{x}_{i, \ell}| = 0$ and $\lim_{t \rightarrow + \infty} |\tilde{v}_{i, \ell}| = 0$. In other words, the Lyapunov functional candidate $V$ is non-increasing and bounded for all time $t \in [0, +\infty]$. 

   If two ABSs satisfy $r_{c} \leq \|\bm{x}_{i}(0) - \bm{x}_{k}(0)\|$ for all $i, k \in \mathcal{P} \cup \mathcal{Q}$, it is obviously that the collision avoidance potential function \eqref{Eq: CA-function-V} satisfies 
	\begin{equation}\label{eq: CA}
		\lim_{\|\bm{x}_{i} - \bm{x}_{j}\| \rightarrow r_{c}^{+}} V^{(i,k)}\left(\bm{x}_{i}, \bm{x}_{k}\right)  = +  \infty. 
	\end{equation}
    Here, $\lim_{x \rightarrow a^{+}} f(x)$ is the right-hand limits of function $f(x)$ at $a$. Based on \eqref{eq: CA}, it can be concluded that collisions between the ABSs can be avoided.

\bibliographystyle{IEEEtran}
\bibliography{reference}

	\begin{IEEEbiography}
		[{\includegraphics[width=1in, height=1.25in, clip, keepaspectratio]{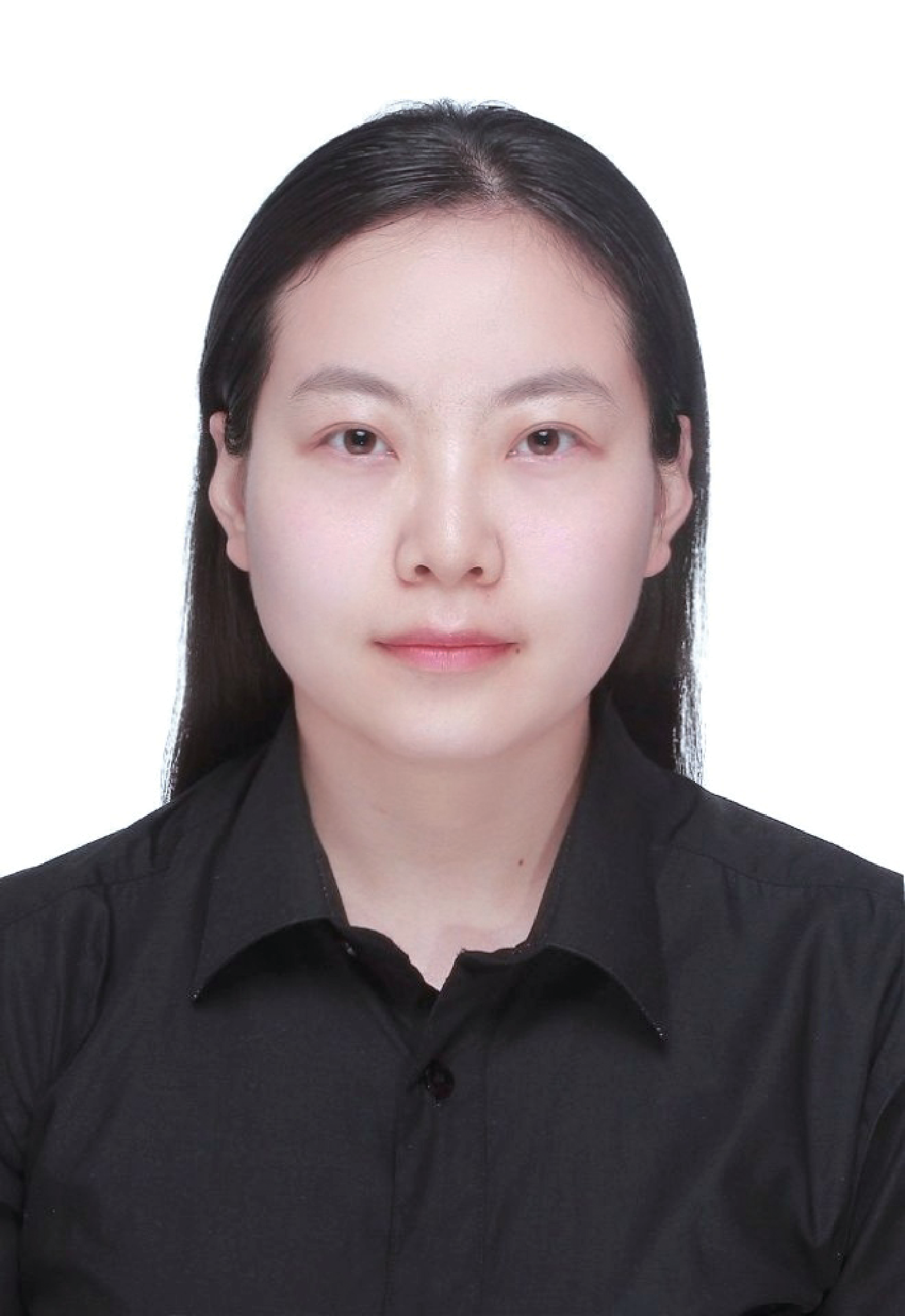}}]{Xiao Fan} received the B.S. and the M.S. degrees in Mathematics from the Shanxi Datong University, Datong, China, in 2015, and the Guilin University of Electronic Technology, Guilin, China, in 2018, respectively. She received the Ph.D. degree in Information and Communication Engineering from Sun Yat-sen University, Guangzhou, China, in 2025. She is now an assistant professor at the Guilin University of Electronic Technology. Her research interests include cooperative UAV communications and nonlinear systems.
	\end{IEEEbiography}

\begin{IEEEbiography}
	[{\includegraphics[width=1in, height=1.25in, clip, keepaspectratio]{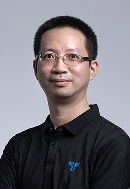}}]{Wenkun Wen} (Member, IEEE) received the Ph.D. degree in Telecommunications and Information Systems from Sun Yat-sen University, Guangzhou, China, in 2007. Since 2020, he has been with Techphant Technologies Co. Ltd., Guangzhou, China, as Chief Engineer.

From 2008 to 2009, he was with the Guangdong-Nortel R\&D center in Guangzhou, China, where he worked as a system engineer for 4G systems. From 2009 to 2012, he worked at the LTE R\&D center of New Postcom Equipment Co. Ltd., Guangzhou, China, where he served as the 4G standard team manager. From 2012 to 2018, he was with the 7th Institute of China Electronic Technology Corporation (CETC) as an expert in wireless communications. From 2018 to 2020, he served as Deputy Director of the 5G Innovation Center at CETC. His research interests include 5G/B5G mobile communications, machine-type communications, narrow-band wireless communications, and signal processing.
\end{IEEEbiography}	

\vfill

\begin{IEEEbiography}
	[{\includegraphics[width=1in, height=1.25in, clip, keepaspectratio]{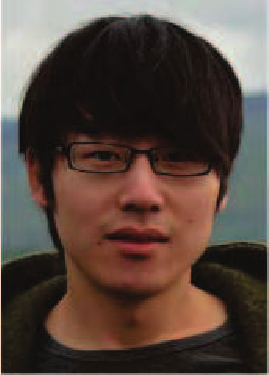}}]{Peiran Wu} (Member, IEEE) received the Ph.D. degree in electrical and computer engineering from The University of British Columbia (UBC), Vancouver, Canada, in 2015.
	
	From October 2015 to December 2016, he was a Post-Doctoral Fellow at UBC. In the Summer of 2014, he was a Visiting Scholar with the Institute for Digital Communications, Friedrich-Alexander-University Erlangen-Nuremberg (FAU), Erlangen, Germany. Since February 2017, he has been with Sun Yat-sen University, Guangzhou, China, where he is currently an Associate Professor. Since 2019, he has been an Adjunct Associate Professor with the Southern Marine Science and Engineering Guangdong Laboratory, Zhuhai, China. His research interests include mobile edge computing, wireless power transfer, and energy-efficient wireless communications. 
\end{IEEEbiography}

\vfill

\begin{IEEEbiography}
	[{\includegraphics[width=1in,height=1.25in,clip,keepaspectratio]{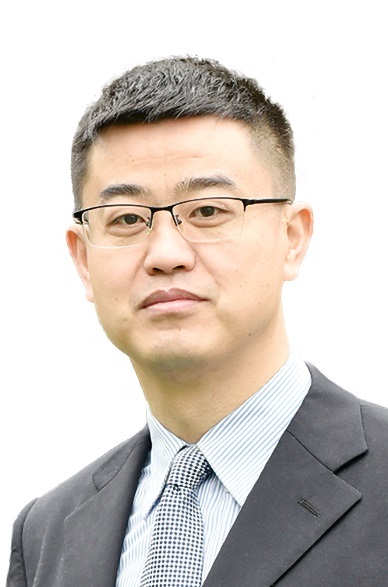}}]{Junhui Zhao} 
	(Senior Member, IEEE) received the M.S. and Ph.D. degrees from Southeast University, Nanjing, China, in 1998 and 2004, respectively. 
	
	From 1998 to 1999, he worked with Nanjing Institute of Engineers at ZTE Corporation. Then, he worked as an Assistant Professor in 2004 at the Faculty of Information Technology, Macao University of Science and Technology, and continued there till 2007 as an Associate Professor. In 2008, he joined Beijing Jiaotong University as an Associate Professor, where he is currently a Professor at the School of Electronics and Information Engineering. Meanwhile, he was also a short-term Visiting Scholar at Yonsei University, South Korea in 2004 and a Visiting Scholar at Nanyang Technological University, Singapore from 2013 to 2014. His current research interests include wireless communication, internet of things, and information processing in traffic.
\end{IEEEbiography}

\vfill
	
\begin{IEEEbiography}
	[{\includegraphics[width=1in, height=1.25in, clip, keepaspectratio]{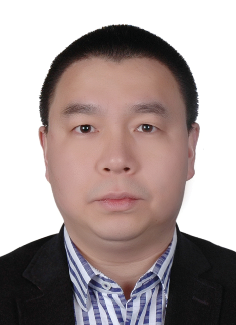}}]{Minghua Xia} (Senior Member, IEEE) received the Ph.D. degree in Telecommunications and Information Systems from Sun Yat-sen University, Guangzhou, China, in 2007.
	
	From 2007 to 2009, he was with the Electronics and Telecommunications Research Institute (ETRI) of South Korea, Beijing R\&D Center, Beijing, China, where he worked as a member and then as a senior member of the engineering staff. From 2010 to 2014, he was in sequence with The University of Hong Kong, Hong Kong, China; King Abdullah University of Science and Technology, Jeddah, Saudi Arabia; and the Institut National de la Recherche Scientifique (INRS), University of Quebec, Montreal, Canada, as a Postdoctoral Fellow. Since 2015, he has been a Professor at Sun Yat-sen University. Since 2019, he has also been an Adjunct Professor with the Southern Marine Science and Engineering Guangdong Laboratory (Zhuhai). His research interests are in the general areas of wireless communications and signal processing.
\end{IEEEbiography}

\end{document}